\documentclass[12pt]{article}
\usepackage{jheppub}
\pdfoutput = 1

\usepackage{amsmath}
\usepackage{amssymb}
\usepackage{bbm}
\usepackage{bbold}
\usepackage{braket}
\usepackage{caption}
\usepackage{color}
    \definecolor{darkgreen}{rgb}{0,0.5,0}
    \definecolor{darkred}{rgb}{0.5,0,0}
    \definecolor{darkblue}{rgb}{0,0,0.6}
    \definecolor{purple}{rgb}{0.4,.2,0.7}

\usepackage[mathcal]{eucal}
\usepackage{float}
\usepackage{graphicx}
\usepackage{hyperref}

\usepackage{mathabx}
\usepackage{mathtools}
\usepackage{pdfsync}
\usepackage{slashed}
\usepackage[normalem]{ulem}
\usepackage{upgreek}
\usepackage{url}

\renewcommand\L{Lichnerowicz }

\usepackage[ragged]{footmisc}
\def\be{\begin{equation}}
\def\ee{\end{equation}}

\renewcommand{\tilde}{\widetilde}

\numberwithin{equation}{section}

\begin{document}

\title{The Canonical Ensemble Reloaded: The Complex-Stability of Euclidean quantum gravity for Black Holes in a Box}
\author[a]{Donald~Marolf,}
\author[b]{Jorge~E.~Santos}

\affiliation[a]{Department of Physics, University of California at Santa Barbara, Santa Barbara, CA 93106, U.S.A.}
\affiliation[b]{Department of Applied Mathematics and Theoretical Physics, University of Cambridge, Wilberforce Road, Cambridge, CB3 0WA, UK}

\emailAdd{marolf@ucsb.edu}
\emailAdd{jss55@cam.ac.uk}

\abstract{
We revisit the stability of black hole saddles for the Euclidean path integral describing the canonical partition function $Z(\beta)$ for gravity inside a spherical reflecting cavity.
The boundary condition at the cavity wall couples the transverse-traceless (TT) and pure-trace modes that are traditionally used to describe fluctuations about Euclidean Schwarzschild black holes in infinite-volume asymptotically flat and asymototically AdS spacetimes.
This coupling obstructs the familiar Gibbons-Hawking-Perry treatment of the conformal factor problem, as Wick rotation of the pure-trace modes would require that the TT modes be rotated as well.  The coupling also leads to complex eigenvalues for the \L operator.  We nevertheless find that the \L operator can be diagonalized in the space of coupled modes.  This observation allows the eigenmodes to  define a natural generalization of the pure-trace Wick-rotation recipe used in infinite volume, with the result that  a mode with eigenvalue $\lambda$ is stable when ${\rm Re}\,\lambda > 0$.  In any cavity, and with any cosmological constant $\Lambda \le 0$,  we show this recipe to reproduce the expectation from black hole thermodynamics that large Euclidean black holes define stable saddles while the saddles defined by small Euclidean black holes are unstable.}

\maketitle

\section{Introduction}

Gibbons and Hawking \cite{Gibbons:1976ue} argued long ago that  gravitational partition functions $Z(\beta)$ are naturally described by Euclidean path integrals.  Such integrals can be evaluated in the semiclassical approximation using saddle points associated with Euclidean black holes. The behavior of the action under small fluctuations indicates stability or instability of these saddles and determines which saddles are relevant to a given computation.

Unfortunately, due to the conformal factor problem, the Euclidean gravitational action is unbounded below.  This prevents one from taking the integral over all Euclidean metrics as a definition of the problem to be studied.  Many authors \cite{Hartle:2020glw,Schleich:1987fm,Mazur:1989by,Marolf:1996gb,Dasgupta:2001ue,Ambjorn:2002gr,Feldbrugge:2017kzv,Feldbrugge:2017fcc,Feldbrugge:2017mbc} have argued that the fundamental definition should instead be made in Lorentz signature, with implications for Euclidean path integrals determined by further careful study.
 While we are deeply sympathetic to this point of view, it has not yet led to a useful recipe to determine the stability of general Euclidean saddles.

As a result, past work has sometimes simply proposed a defining contour in particular contexts and then checked that it yields physically sensible results.  In particular, for path integrals that compute partition functions in the canonical ensemble in asymptotically flat or asymptotically AdS spacetimes,  it was suggested in \cite{Gibbons:1978ac} that the conformal factor should be integrated over a contour parallel to the imaginary axis while the conformal metric should be integrated over real values.  
This recipe has a mathematical elegance and has the important property that black holes saddles for the canonical partition function are unstable when they have negative specific heat and when the above classes of perturbations can be decoupled (see
\cite{Prestidge:1999uq} for the asymptotically AdS context and section 4.3 of \cite{Dias:2010eu} for a more general argument; in the latter case it is generally clear that the associated negative-action deformation preserves the determinant of the metric and thus remains a physical negative mode after the above Wick rotation).  

At the level of linearized fluctuations about saddles, this recipe is easiest to implement by first decomposing the Euclidean-signature perturbations into pure-trace, transverse-traceless (TT), and pure-gauge modes.  The proposal then implies that one should integrate the Euclidean pure-trace modes over {\it imaginary} field values while integrating transverse-traceless modes over the real contour.  Since pure-gauge modes do not change the action, the particular contour chosen for such modes will not affect the result and can be chosen to help preserve boundary conditions not discussed in detail in \cite{Gibbons:1978ac}.   

Mode stability in this approach is often discussed by noting (see e.g. \cite{Headrick:2006ti}) that the quadratic action for perturbations $h$ about a saddle $\hat{g}_{ab}$ may be written as an expectation value
\begin{subequations}
\begin{equation}
\label{eq:quadact}
S^{(2)}[h] =  (h,Lh)
\end{equation}
where
\begin{equation}
(h, \tilde h) = \frac{1}{32 \pi G}\int_{\mathcal{M}}\mathrm{d}^d x\,\sqrt{\hat{g}}\,h_{ab}\,\hat{\mathcal{G}}^{ab\;cd} \tilde h_{ab}
\label{eq:DWIP}
\end{equation}
is the inner product defined by the DeWitt$_{-1}$ metric\footnote{It was pointed out by DeWitt \cite{DeWitt:1967yk} that there is a one-parameter family of `ultralocal' metrics
$$\hat{\mathcal{G}}^{ab\;cd}_{\lambda_{DW}}=\frac{1}{2}\left(\hat{g}^{ac}\hat{g}^{bd}+\hat{g}^{ad}\hat{g}^{bc} + \lambda_{DW} \hat{g}^{ab}\hat{g}^{cd}\right)\,$$
on the space of Riemannian metrics on $\mathcal{M}$. Here we have added the subscript $DW$ to DeWitt's parameter $\lambda$ to avoid confusion with the eigenvalues we study below.  Eq. \eqref{eq:DW-1} corresponds to $\lambda_{DW}=-1$, or $a=-1/2$ in the conventions of \cite{Headrick:2006ti}. We refer to the general such metric as DeWitt$_{\lambda_{DW}}$, so \eqref{eq:DW-1} is DeWitt$_{-1}$.}
\begin{equation}
\label{eq:DW-1}
 \hat{\mathcal{G}}^{ab\;cd}=\frac{1}{2}\left(\hat{g}^{ac}\hat{g}^{bd}+\hat{g}^{ad}\hat{g}^{bc}-\hat{g}^{ab}\hat{g}^{cd}\right)\, ,
\end{equation}
$G$ is Newton's constant and
\begin{equation}
(L h)_{ab} = (\hat{\Delta}_L h)_{ab}+2 \hat{\nabla}_{(a}\hat{\nabla}^p \bar{h}_{b)p}
\label{eq:actionungaugedL}
\end{equation}
where
\begin{equation}
\bar{h}_{ab}=h_{ab}-\frac{\hat{g}_{ab}}{2}h\, ,
\end{equation}
and
\begin{equation}
(\hat{\Delta}_L h)_{ab}=-\hat{\nabla}_p \hat{\nabla}^p h_{ab}-2\,\hat{R}_{acbd}\,h^{cd}\,
\end{equation}
\end{subequations}
is the \L operator. In particular, $L$ reduces to the usual \L operator $\hat{\Delta}_L$ on perturbations that satisfy the de Donder gauge, \emph{i.e.} $\hat{\nabla}^a \bar{h}_{ab}=0$. The inner product \eqref{eq:DW-1} has indefinite signature but, in the above thermodynamic context, Wick-rotating the pure-trace modes makes the inner product positive definite while leaving $L$ self-adjoint.  Positivity of the Wick-rotated version of \eqref{eq:quadact} is thus determined by the spectrum of $L$, with positive eigenvalues of $L$ corresponding to stable modes.
In the presence of matter, the pure-trace modes generally couple to matter modes, but by using the approach of \cite{Kol:2006ga} analogous coupled modes can often be identified by inspection with correspondingly successful results \cite{Monteiro:2008wr}. See also \cite{Gratton:1999ya,Gratton:2000fj,Gratton:2001gw} in the cosmological context.

We will be interested below in a more complicated situation that occurs when one studies gravitational thermodynamics in a finite-sized cavity with fixed boundary metric.  As in the case with matter fields, such boundary conditions again couple the pure-trace modes to other modes, which in this case are TT gravitational modes.  This coupling is described in detail in section \ref{sec:intropert}, but it should not be surprising: when they are both real, there are particular combinations of pure-trace perturbations and TT perturbations that preserve the induced metric on the cavity wall, even though each piece separately would change the induced metric. But if we Wick rotate the pure trace modes to imaginary values while keeping the TT modes real, then such cancellations cannot occur and the contributions of pure-trace and TT modes to the induced metric would have to vanish independently.  Naively Wick rotating the pure-trace modes would thus have the effect of imposing an additional non-physical boundary condition which should be expected to invalidate any results.  And since the bulk action is of course the same as in the infinite volume case, the issue is not immediately resolved by following \cite{Kol:2006ga}. 

It thus seems that the contour rotation prescription of \cite{Gibbons:1978ac} must be modified inside a reflecting cavity.  
As a first step in doing so we note that the DeWitt$_{-1}$ metric $\hat{\mathcal{G}}^{ab\;cd}$ can still be used to define an operator $L$ via \eqref{eq:quadact}. We study this $L$ below and find that it can again be diagonalized, though some of its eigenvalues may be complex\footnote{\label{foot:BC1}There are active discussions as to whether the full non-linear theory is physically sensible with such Dirichlet boundary conditions; see \cite{Anderson2008,Anderson:2007jpe,Witten:2018lgb,Fournodavlos:2020wde,Fournodavlos:2021eye} for discussions of mathematical issues and  \cite{Andrade:2015qea} for discussion of more physical issues. However the linearized problem is well-defined in the context we study below.  Furthermore, we have verified that following \cite{Anderson:2007jpe,Adam:2011dn,Figueras:2011va,Witten:2018lgb} and using instead boundary conditions that fix the conformal metric and the trace of the extrinsic curvature again leads both to couplings between the pure-trace and TT modes and to complex eigenvalues for the \L operator, though we leave a detailed study of such boundary conditions future work.}.
We then use this observation to propose that the pure-trace Wick-rotation  recipe from infinite volume be generalized to Wick-rotation of appropriate negative-norm parts of the $L$ eigenmodes.   The upshot of our generalization turns out to be that mode stability is again determined by the eigenvalues $\lambda$ of $L$, with stable modes having ${\rm Re}\,\lambda >0$.  This is in turn equivalent to analyzing stability under Ricci flow (perhaps modified by a cosmological term) as proposed in \cite{Headrick:2006ti}, though that reference did not explicitly discuss either complex eigenvalues or a definite proposal for Wick-rotation.

In either our formulation or that of \cite{Headrick:2006ti}, the critical input lies in giving a preferred status to the DeWitt$_{-1}$ inner product.  From our present point of view, this choice is as \emph{ad hoc} as the Wick-rotation of pure-trace modes in \cite{Gibbons:1976ue,Gibbons:1978ji}.  Nonetheless, its physical viability is demonstrated below by showing (again, in analogy with \cite{Prestidge:1999uq}) that the above recipe reproduces the prediction from black-hole thermodynamics that in any cavity the large and small black hole saddles are respectively stable and unstable.

We begin with general formalism in section \ref{sec:formalism}, describing both the sense in which complex eigenvalues are natural and our proposed Wick-rotation.  This section also introduces the particular approach we will take to discretizing our system for numerical investigation, which involves discretizing the system at the level of the action (from which a discrete $L$ operator follows immediately) rather than using the continuum action to define a continuum $L$ (and then attempting to discretize the result).  This discussion also refers to some useful properties of the de Donder gauge that are reviewed in appendix \ref{sec:DDgauge}.

 This sets the scene for section \ref{sec:SAdS} to study fluctuations around the associated Euclidean black holes in $d=4,5$ spacetime dimensions.  In particular, we show there that our proposed Wick-rotation reproduces the thermodynamic stability/instability of large/small black holes.  The associated thermodynamic calculations are of a standard form but, since they do not seem to appear in the existing literature for black holes in a reflecting cavity with negative cosmological constant, we provide the details of this analysis in appendix \ref{sec:AdStherm}.  We close with some final comments and discussion in section \ref{sec:disc}.

\section{Formalism: Wick rotation, stability,  and discretization}
\label{sec:formalism}


\subsection{Wick Rotation and mode stability}
\label{sec:Wick}

It is useful to begin by stating our Wick rotation prescription in very general terms.   We thus consider any action\footnote{The reason for the check ($\check{}$) will be explained in section \ref{sec:further}. } $\check{\mathbb{S}}$, which is a quadratic function of independent and unconstrained real field variables ${\mathbb{Q}}^I$ that is stationary at $\mathbb{Q}^I=0$, and where we will explain the reason for the decoration $\check{}$ in section \ref{sec:further}.  Setting $\check{\mathbb{S}}=0$ at the stationary point we thus have
\begin{equation}
\label{eq:genact}
\check{\mathbb{S}} = \sum_{IJ} {\mathbb{Q}}^I \check{\mathbb{S}}_{,IJ} {\mathbb{Q}}^J,
\end{equation}
where the subscript $,IJ$ denotes the indicated second derivative of the action.  We take our action to be real for real $\mathbb{Q}^I$, so that the coefficients $\check{\mathbb{S}}_{,IJ}$ are real as well.
Since our discussion is general, the label $I$ may take either continuous values (as in the problem we wish to study) or discrete values (as in the numerical approximations that we will use in practice).  In the latter case $\sum_I, \sum_{IJ}$ represents the appropriate integral and $,IJ$ denotes functional derivatives.  The above notation is chosen to match that used in later sections in our study of particular systems.

Our goal will be to investigate positivity of \eqref{eq:genact}.  However, since we will study Euclidean gravity, the conformal factor problem will ensure that \eqref{eq:genact} is not positive definite for real ${\mathbb{Q}}^I$.    We therefore wish to introduce some structure that defines a `Wick rotation' ${\mathbb{Q}}^I = \sum_J W^I_J {\mathbb{R}}^J$ for some complex matrix $W^I_J$ and to instead investigate positivity of $\check{\mathbb{S}}$ for real ${\mathbb{R}}$ using

\begin{equation}
\label{eq:genact2}
\check{\mathbb{S}} = \sum_{IJKM} {\mathbb{R}}^I  W^I_M \check{\mathbb{S}}_{,IJ} W^J_K  {\mathbb{R}}^J.
\end{equation}

As foreshadowed in the introduction, we will define our Wick rotation $W^I_J$ by making use of a (real) metric ${\mathbb{G}}_{IJ}$ on the configuration space, which in practice will be the DeWitt$_{-1}$ metric or a discrete approximation thereof.  Indeed, one property we require (which in some sense justifies the name `Wick rotation') is that the `Wick rotated metric'
\begin{equation}
\label{eq:wrm}
\mathbb{N}_{IJ}\equiv\sum_{KL} W^K_I {\mathbb{G}}_{KL} W^L_J
\end{equation}
be positive definite.  We also require $W^I_J$ to be invertible so that the Wick rotation can be undone.

However, this property alone does not suffice to define a unique $W^I_J$.  To specify a particular Wick rotation, we first use the inverse metric ${\mathbb{G}}^{IJ}$ to construct a linear operator
\begin{equation}
\label{eq:linop}
{\mathbb{L}}_J^I = \sum_K {\mathbb{G}}^{IK} \check{\mathbb{S}}_{,KJ}.
\end{equation}
It is straightforward to show that \eqref{eq:linop} is always self adjoint with respect to ${\mathbb{G}}_{IJ}$. That is to say, if we define 
\begin{equation}
\label{eq:ipdef}
(\alpha, \beta) \equiv \sum_{IJ} \alpha_I^* \mathbb{G}_{IJ} \beta_J
\end{equation}
where $*$ denotes complex conjugation, and if we write $(\mathbb{L}\beta)_I =\mathbb{L}_I^J \beta_J$, we have
\begin{equation}
\label{eq:selfadj}
(\mathbb{L} \alpha, \beta) = (\alpha, \mathbb{L} \beta) \ {\rm for\  all} \  \alpha_I, \beta_J.
\end{equation}

We now assume that ${\mathbb{L}}_J^I$ can be diagonalized; \emph{i.e.}, that it's eigenvectors space the space of all $\mathbb{Q}^I$.  Since $\mathbb{G}_{IJ}$ is not positive definite, this does {\it not} follow from the self-adjointness property \eqref{eq:selfadj}.  Instead, it is a property that must be checked for some particular choice of action $\check{\mathbb{S}}$ and metric $\mathbb{G}_{IJ}$.  We will find below that this property does indeed hold in the systems we study for the linearized Einstein-Hilbert action and the DeWitt$_{-1}$ metric\footnote{It should be noted that generic finite-dimensional matrices can in fact be diagonalized in the above sense.  This follows from the fact that generic $n$-dimensional matrices $A$ have $n$ distinct roots of the characteristic equation $det(A-\lambda \mathbb{1}) =0$ and that each distinct root yields a linearly independent eigenvector. As a result, this property is non-trivial only at special points in the parameter space where otherwise-distinct eigenvalues become degenerate.}.

The indefinite signature of $\mathbb{G}_{IJ}$  also allows the eigenvalues $\lambda$ of ${\mathbb{L}}_J^I$ to be complex\footnote{As a simple example, consider the action $\check{\mathbb{S}} = \frac{1}{2}(\mathbb{Q}^1)^2 + 2(\mathbb{Q}^1)(\mathbb{Q}^2) + \frac{3}{2} (\mathbb{Q}^2)^2$ and the line element $ds^2 = \mathbb{G}_{IJ} \mathbb{Q}^I \mathbb{Q}^J = (\mathbb{Q}^1)^2 + 3(\mathbb{Q}^1)(\mathbb{Q}^2) + \frac{1}{2} (\mathbb{Q}^2)^2$.
It is a simple exercise to show that the eigenvalues of \eqref{eq:linop} are $\lambda=\frac{5}{14}\pm\frac{i \sqrt{3}}{14}$, which are indeed complex.}.  In particular, following the standard argument one finds that for eigenvectors $v_1,v_2$ with eigenvalues $\lambda_1, \lambda_2$ we have

\begin{equation}
\lambda_1^* (v_1,v_2) = (\mathbb{L} v_1, v_2) = (v_1,\mathbb{L} v_2) = \lambda_2(v_1,v_2).
\end{equation}
This requires 
\begin{equation}
\label{eq:eigorth}
(v_1,v_2)=0 \ \ {\rm for} \ \  \lambda_1 \neq \lambda_2^*.
\end{equation}
 In particular, taking $v_1=v_2$ requires $\lambda_1 = \lambda_2$ to be real when $(v_1,v_1)\neq 0$, but allows complex eigenvalues for eigenvectors with norm zero.  Note that since the inner product is non-degenerate, for diagonalizable  ${\mathbb{L}}_J^I$ any complex eigenvalues must in fact appear in complex-conjugate pairs $\lambda, \lambda^*$.

In our case this last property is also manifest from the fact that both $\mathbb{G}_{IJ}$ and $\check{\mathbb{S}}_{,IJ}$ are real so that $\mathbb{L}_J^I$ is also real.  Thus if $\sum_J \mathbb{L}_J^I v^J = \lambda v^I$, then complex-conjugating this result yields $\sum_J \mathbb{L}_J^I (v^J)^* = \lambda^* (v^I)^*$.  In particular, complex eigenvalues require complex eigenvectors $v, v^*$ from which we can form the real linear combinations ${\rm Re}(v) = \frac{v+v^*}{2}$ and ${\rm Im}(v) = \frac{v-v^*}{2i}$.

When the eigenvalues of $\mathbb{L}_J^I$ are non-degenerate, our Wick rotation will be defined to act in a simple way on the associated eigenspaces.  It is convenient to state this definition by noting that any eigenvalue $\lambda$ can be associated with an eigenvector $v$ that satisfies the normalization condition
\begin{equation}
\label{eq:norm}
(v,v^*)=1 = (v^*,v).
\end{equation}
Indeed, given an arbitrary eigenvector $V$ we may define $\alpha = \sqrt{(V^*,V))}$ so that the rescaled eigenvector $v=\alpha^{-1} V$ satisfies \eqref{eq:norm}.  Here we use the fact that combining the non-degeneracy of the spectrum with the non-degeneracy of the inner product requires $\alpha$ to be non-zero.  

When $\lambda$ is real, $v^*$ corresponds to the same eigenvalue, so non-degeneracy of the spectrum and Hermiticity of the inner product requires $v^*$ to be $v$ times a phase.  The condition \eqref{eq:norm} then also requires $(v,v)=\pm 1$.  This sign is not a convention, but a statement of whether the given eigenmode has positive or negative norm.

The convention \eqref{eq:norm} gives a certain preferred status to the real and imaginary parts ${\rm Re}(v)$, ${\rm Im}(v)$.  In particular, for $\lambda \neq \lambda^*$ one finds the following inner products:
\begin{subequations}
\begin{align}
&({\rm Re}(v),{\rm Re}(v))= \frac{1}{2}\,,
\\
&({\rm Im}(v),{\rm Im}(v))= -\frac{1}{2}\,,
\\
&({\rm Re}(v),{\rm Im}(v))= 0.
\end{align}
\label{eq:RIip}
\end{subequations}
Note that more generally one finds $({\rm Re}(v),{\rm Im}(v)) = \frac{1}{2}{\rm Im}(v^*,v)$. Up to a choice of overall scale  we thus see that requiring orthogonality of ${\rm Re}(v)$, ${\rm Im}(v)$ is equivalent to imposing $(v,v^*)=\pm 1$, and choosing the positive sign imposes the convention that  ${\rm Re}(v)$ is the positive-norm member of the pair ${\rm Re}(v)$, ${\rm Im}(v)$ while ${\rm Im}(v)$  is the negative norm member.

In this context we define $W_J^I$ by the following properties:
\begin{enumerate}
\item $W_J^I$ leaves invariant any eigenvector with real eigenvalue $\lambda$ and positive norm $(v,v)$.

\item Any eigenvector with real eigenvalue $\lambda$ and negative norm $(v,v)$ is an eigenvector of $W_J^I$ with eigenvalue $i$.

\item For complex eigenvalues $\lambda$ (with $\lambda \neq \lambda^*$), the corresponding ${\rm Re}(v)$ left invariant by $W_J^I$ while ${\rm Re}(v)$ is an eigenvector of $W_J^I$ with eigenvalue $i$.
\end{enumerate}
From \eqref{eq:RIip} it is manifest that  \eqref{eq:wrm} is then positive-definite as desired.  But of course the above proposal is far from unique in this regard.  In particular,  as noted in the introduction, from our present perpective the proposal is as \emph{ad hoc} as the Wick rotation of the conformal factor proposed in \cite{Gibbons:1978ac}.  As a reminder of this, we will refer to the above proposal and to the further refinements below as a `rule of thumb.'

Let us ignore for the moment any issues associated with gauge invariance of the gravitational action, returning to such issues in section \ref{sec:further} below.  Non-degeneracy of the spectrum of $\mathbb{L}^I_J$ is then generic, and it remains to define $W_J^I$  only at those special points in parameter space where eigenvalues become degenerate.  In such cases we simply define $W_J^I$ by requiring our Wick rotation to be a continuous function of the parameters.  We hypothesize that this is always possible, and we will in section \ref{sec:results} below that this property holds for the particular problem studied here.

With this prescription it is clear that positivity of \eqref{eq:genact2} is determined by the spectrum of $\mathbb{L}_J^I$.  In particular, since the eigenmodes $v_i$ of $\mathbb{L}_J^I$ span the space, we may write any $\mathbb{R}^I$ as a linear combination of such modes.  Due to \eqref{eq:eigorth}, it will be useful to divide the full set of eigenvalues $v_i$ into the real eigenvalues $\lambda_a$ with eigenvectors $V_a$ (chosen to have all components real), the eigenvalues $\lambda_A$ with positive imaginary parts with eigenvectors $v_A$, and the complex-conjugate eigenvalues $\lambda_A^*$ with eigenvectors $v_A^*$.  We may then write
$\mathbb{R}^I = \sum_a c^a V_a^{I} + \sum_A (C^A v_A^I + C^{A*} v_A^{I*}) $ with real $c^a$ and use \eqref{eq:linop}, \eqref{eq:eigorth}, and the definition of $W^I_J$ to find the final Wick-rotated result 
\begin{equation}
\label{eq:genact3}
\check{\mathbb{S}} = \sum_{a}  \lambda_a (c^a)^2 + 2 \sum_A ({\rm Re}\;\lambda_A) |C^A|^2.
\end{equation}
Since the $c^a$ are real, it is clear from \eqref{eq:genact3} that a given mode is stable when it satisfies ${\rm Re}\;\lambda > 0$, with marginal stability for ${\rm Re}\;\lambda = 0$ and strict instability for ${\rm Re}\;\lambda < 0$. We will investigate this definition of stability for gravity in a box in section \eqref{sec:SAdS}, where we also discuss the relationship to the Ricci-flow prescription of \cite{Headrick:2006ti}.

\subsection{Further issues and a comment on numerics}
\label{sec:further}

Our discussion above was cleanly presented in terms of an action $\check{\mathbb{S}}$ that was a quadratic function of independent and unconstrained field variables $\mathbb{Q}^I$.  However, the form in which gravitational systems are typically presented is not quite so clean.  One issue is that, because spacetime is continuous, a good gravitational variational principle will generally require certain boundary conditions that constrain what would otherwise be independent values of the $\mathbb{Q}^I$.  At first glance this is a minor issue that should affect only our choice of notation.  We use the somewhat awkward notation $\tilde{\mathbb{Q}}^{\tilde I}$ to denote a naive set of field variables on which boundary conditions have not yet been imposed, and we reserve $\mathbb{Q}^I$ for the remaining independent variables after the constraints defined by such boundary conditions have been solved.  Since we study linearized theories, the boundary conditions are also linear and explicit such solutions are always possible.

However, due to the need to impose boundary conditions, and in order to develop a practical formalism that can be applied to very general situations, when it comes time to study our system numerically we will find it prudent to depart somewhat from the traditional approach of \cite{Gibbons:1978ji,Gross:1982cv,Allen:1984bp,Prestidge:1999uq,Headrick:2006ti,Monteiro:2009ke,Monteiro:2009tc,Dias:2009iu,Dias:2010eu}.  That approach would use the continuum action to construct a continuum differential operator $\mathbb{L}$, after which the boundary conditions could be used to define an appropriate discretization of this operator.  However, this discretization is not unique, and may not interact cleanly with the discretization of other structures such as the metric $\mathbb{G}$.  

For this reason, we instead choose to first discretize the fields $\mathbb{Q}$ that solve the boundary conditions, the action $\check{\mathbb{S}}$, and the chosen metric $\mathbb{G}$.  Using the discretized $\check{\mathbb{S}},\mathbb{G}$ to define $\mathbb{L}$ via \eqref{eq:linop} then guarantees that the properties described in section \ref{sec:Wick} hold exactly, or at least up to the numerical accuracy with which the resulting algebraic equations have been solved.  While this is a minor issue for the current paper, we have found it to be extremely useful in studying the more general boundary conditions that we will consider in \cite{MicroPaper}. This is the comment on our numerical approach advertised in the title of this subsection

A second issue is that gravitational systems have a gauge symmetry, so that different $\mathbb{Q}^I$ are generally not physically independent.  While such physical independence was not strictly required in our treatment above, its lack means that -- even after discretization -- general physical parameters might not lead to non-degeracy of the spectrum of $\mathbb{L}$.    Indeed, gauge invariance means that there are modes $\xi^I$ which we may use to shift any $\mathbb{Q}^I$ without changing the value of the action; \emph{i.e.}, the action is the same for all $\mathbb{Q}^I + \epsilon \xi^I$, independent of the value of $\epsilon$.  For gravity we may take $\xi^I$ real, so that for real $\mathbb{Q}^I$ we may differentiate the above result with respect to $\epsilon$ to find $(\mathbb{Q}, \mathbb{L} \xi) =0$ for all $\mathbb{Q}$.  And since the inner product is non-degenerate, this would require each pure gauge mode $\xi^I$ to be a zero-eigenvalue eigenvector of $\mathbb{L},$ leading to a highly degenerate spectrum. 

While this is rightly considered a mere technical issue, it is one that we will wish to avoid. As usual, one can do so by fixing a gauge.  The de Donder gauge $\nabla^a h_{ab} - \frac{1}{2} \nabla_b h =0$ is a common choice.  At the conceptual level we will also make this choice, though with certain technical modifications as discussed below.

As reviewed in appendix \ref{sec:DDgauge}, the time-component of the de Donder gauge condition is straightforward to implement in our context.  Indeed, doing so corresponds merely to setting a certain component of the metric to zero.  But the radial component will also be non-trivial in our context.  While it can also be solved, inserting the solution into the action leads to a higher-derivative action that is more complicated to study.

We thus avoid imposing this last condition explicitly (except at the cavity wall where it use it as an additional boundary condition).   Instead, we will call the quadratic gravitational action ${\mathbb{I}}$ (with no check ($\check{}$)), and we will define a new action 
\begin{equation}
\label{eq:Smod}
\check{\mathbb{I}}(\mathbb{Q}) \equiv {\mathbb{I}}(\mathbb{Q}) + \sum_{IJ} \mathbb{K}_{IJ} \mathbb{Q}^I\mathbb{Q}^J.
\end{equation}
 to be used in our numerics by adding an additional quadratic term $\sum_{IJ} \mathbb{K}_{IJ} \mathbb{Q}^I\mathbb{Q}^J$ that explicitly breaks gauge invariance but which has no effect on modes that satisfy the de Donder condition.  That is to say, we require $\sum_{IJ} \mathbb{K}_{I} \mathbb{Q}^I =0$ when $\mathbb{Q}^I$ satisfies the de Donder condition. In our gravitational problem, it is the operator $\mathbb{L}$ defined by the modified action \eqref{eq:Smod} that turns out to be usual \L operator in the bulk.  
 
 While one could in principle use any gauge condition in the above way, the de Donder gauge has a particularly elegant property for this purpose.  To explain this properly, let us use $V$ to denote the space of all pure-gauge modes.  Then as also reviewed in appendix \ref{sec:DDgauge}, as defined by the DeWitt$_{-1}$ inner product, in our context the orthogonal complement $V^\perp$ of $V$ consists precisely of perturbations satisfying the de Donder gauge.  To see the utility of this property, recall that following the strategy outlined in section \ref{sec:Wick} means that  we study eigenvalues of a self-adjoint operator $\mathbb{L}$. And since the action of this operator on any de Donder gauge mode is the same as that defined by gauge-fixing the theory, this $\mathbb{L}$ must have eigenmodes that satisfy the de Donder gauge and which in fact span $V^\perp$.  The above orthogonality then means that any eigenmode will either lie in $V^\perp$ or in $V$; \emph{i.e.}, the eigenmodes will sort themselves cleanly into "physical" de Donder gauge modes and pure-gauge modes. 
 
In particular, recall that (as reviewed in appendix \ref{sec:DDgauge}) any mode can be uniquely decomposed into a pure-gauge mode and a mode that satisfies the de Donder gauge. Since the addition of a pure-gauge mode leaves the original ${\mathbb{S}}$ invariant, if this action has a negative mode then it must in fact have a negative mode that satisfies the de Donder gauge\footnote{The same is true of zero modes, though one should understand that Killing fields of the background give rise to trivial zero modes for which the metric perturbation $h_{ab}$ vanishes identically.  Since the isometries of Euclidean AdS-Schwarzschild form a compact group, such zero modes cannot destabilize our canonical ensemble.}.  But since the two actions agree on such modes, this would also provide a negative mode for the modified action $\check{\mathbb{S}}$.  It follows that establishing positivity of $\check{\mathbb{S}}$ will also establish positivity of the original action $\mathbb{S}$ under physical perturbations.

With these details in hand, we are now ready to analyze our gravitational system.

\section{Fluctuations about the $d=4,5$ Euclidean Schwarzschild(-AdS) black hole in a reflecting cavity}
\label{sec:SAdS}
We now turn to the details of mode stability for Euclidean  Schwarzschild-AdS black holes in reflecting cavities. We introduce notation and conventions for the metric and for time-independent spherical perturbations in section \ref{sec:intropert} and provide brief comments on numerical methods in section \ref{sec:numerics}. Results regarding eigenvalues and eigenvectors are then presented in section \ref{sec:results}.

\subsection{Linearized modes}
\label{sec:intropert}

Euclidean AdS-Schwarzschld black hole solutions in a spherical cavity are just those parts of the usual  Euclidean  Schwarzschild-AdS  spacetimes that lie within the cavity walls. We will work in Schwarzschild coordinates, where the Euclidean Schwarzschild-AdS line element reads
\begin{equation}
\label{eq:SAdSmetric}
\mathrm{d}s^2=f(r)\,\mathrm{d}\tau^2+\frac{\mathrm{d}r^2}{f(r)}+r^2\,\mathrm{d}\Omega_{d-2}^2\,,
\end{equation}
with $\mathrm{d}\Omega_n^2$ the metric on a unit radius round $n-$sphere, with
\begin{equation}
f(r)=\frac{r^2}{\ell^2}+1-\left(\frac{r_+}{r}\right)^{d-3}\left(\frac{r_+^2}{\ell^2}+1\right)\,.
\end{equation}
and where $\ell$ is the AdS length scale, related to the cosmological constant $\Lambda$ via
\begin{equation}
\Lambda = -\frac{(d-1)(d-2)}{2\ell^2}\,.
\end{equation}

The black hole event horizon is located at $r=r_+$, where $f(r)$ vanishes linearly. We take  $\tau$ to be a periodic coordinate with $\tau\sim \tau+\beta_*$, and to avoid a conical singularity at $r=r_+$, we must have
\begin{equation}
\beta_* = \frac{4\pi}{\left|f^\prime(r_+)\right|}=\frac{4 \ell^2 \pi  r_+}{(d-3) \ell^2+(d-1) r_+^2}\, .
\label{eq:cav}
\end{equation}
This fixes $r_+$ as a function of $\beta_*$. Note that $\beta_*$ is {\it not} the physical inverse temperature as measured at the cavity wall, as such walls have not yet been introduced. But in any case we will find it more convenient to describe results in terms of $r_+$ below.  

The detailed thermodynamics of such solutions in a cavity of radius $r_0$ are described in appendix \ref{sec:AdStherm}.  As usual, for a given box size there are both `large' and `small' branches of the space of black hole solutions, with the large/small black holes being thermodynamically stable/unstable and with the boundary between the two branches given by \eqref{eq:crazyl}. The calculations are standard, and we will use \eqref{eq:crazyl} below. 

Since the background metric is spherically symmetric, we can take advantage of its $SO(d-1)$ symmetry group and expand perturbations in terms of spherical harmonics.  These harmonics come in three classes: scalar-derived gravitational perturbations, vector-derived gravitational perturbations and tensor-derived gravitational perturbations, with all of these being mutually orthogonal with respect to any DeWitt inner product.  Furthermore, we expect modes with non-zero angular momentum to have larger values of $\mathrm{Re}\,\lambda$, so we will study in detail only perturbations with zero angular momentum.

For zero angular momentum the vector-derived and tensor-derived modes are pure gauge and thus leave the action invariant.  When the cavity walls taken to infinity, adding angular momentum is known to increase the eigenvalues of the \L operator \cite{Kudoh:2006bp}. 
Assuming that this result continues to hold in the presence of a finite-radius spherical wall, all vector- and tensor-derived modes will necessarily have positive Euclidean action  after applying the Wick rotation of section \ref{sec:Wick}.  We reserve a careful check of this assumption for future work.

We are thus left with scalar perturbations. Scalar perturbations depend on a single quantum number $\ell_S\geq0$ and, as described above, modes with $\ell_S=0$ will have the smallest $\mathrm{Re}\,\lambda$ eigenvalue $\lambda$.  We will therefore focus on such modes.

Recalling that we restrict  discussion to time-independent geometries, these $\ell_S=0$ perturbations have exactly the same symmetries as the background geometry. Furthermore, as reviewed in appendix \ref{sec:DDgauge}, imposing the $\tau$-component of the de Donder gauge condition sets the $r\tau$ component of the perturbation to zero everywhere. As a result, our perturbations take the form
\begin{equation}
\delta \mathrm{d}s^2 = a(r)f(r)\mathrm{d}\tau^2+\frac{b(r)}{f(r)}\mathrm{d}r^2+c(r)\,r^2 \mathrm{d}\Omega_{d-2}^2\, .
\label{eq:lineSO3}
\end{equation}

The action (\ref{eq:quadact}) is readily evaluated on such perturbations.  However, as described in section \ref{sec:further} we in fact wish to modify this action by adding a term that breaks gauge invariance.  In particular, a standard argument shows that using the DeWitt$_{-1}$ metric the space $V'{}^\perp$ described in section \ref{sec:further} is just the space of perturbations that satisfy the de Donder gauge.  As a result, we may take the additional $\mathbb{K}$ term in \eqref{eq:Smod} to be given by $(h,\tilde h)$ with $\tilde h_{ab} = - 2 \hat{\nabla}_{(a}\hat{\nabla}^p \bar{h}_{b)p}$, so that the modified action is just 
\begin{eqnarray}
\check{S}^{(2)} &=& (h, \hat \Delta_L h)\nonumber
\\
&=& \frac{\varepsilon^2}{32 \pi G}\int_{\mathcal{M}}\mathrm{d}^d x\,\sqrt{\hat{g}}\,h_{ab}\,\hat{\mathcal{G}}^{ab\;cd}(\hat{\Delta}_L h)_{cd}\,.
\label{eq:actiongauged}
\end{eqnarray}
evaluated on the perturbed line element (\ref{eq:lineSO3}). The results may be written
\begin{equation}
\check{S}^{(2)} =\frac{\varepsilon^2 \Omega_{d-2}}{64 \pi G} \int_{r_+}^{r_0} \mathrm{d}r\,r^{d-2}\vec{q}\cdot\left[\frac{1}{r^{d-2}}\mathbf{P}\cdot \frac{\mathrm{d}}{\mathrm{d}r} \left(f\,r^{d-2} \frac{\mathrm{d}\vec{q}}{\mathrm{d}r}\right)+\mathbf{V}\cdot \vec{q}\right]\,,
\label{eq:actionneg}
\end{equation}
where $\Omega_{d-2}$ is the volume of the metric on a unit radius round $(d-2)-$sphere, $\vec{q}=\{a(r),b(r),c(r)\}$, the operation $\cdot$ denotes the standard Cartesian inner product in Euclidean space defined by the Kronecker delta metric $\delta_{\tilde I, \tilde J}$, and $\mathbf{P}$ and $\mathbf{V}$ are symmetric matrices with the following independent components
\begin{equation}
\mathbf{P}_{11}=\mathbf{P}_{22}=-1\,, \quad \mathbf{P}_{12}=1\,,\quad  \mathbf{P}_{13}=\mathbf{P}_{23}=d-2\,,\quad\text{and}\quad  \mathbf{P}_{33}=(d-4)(d-2)\,.
\end{equation}
and
\begin{align}
&\mathbf{V}_{11}=-\frac{(d-2) f'(r)}{r}+\frac{f'(r)^2}{f(r)}-f''(r) \nonumber
\\
& \mathbf{V}_{12}=-\frac{(d-2) f'(r)}{r}-\frac{f'(r)^2}{f(r)}+f''(r)\nonumber
\\
& \mathbf{V}_{13}=\frac{(d-3) (d-2)}{r^2}-\frac{(d-3) (d-2) f(r)}{r^2}-\frac{(d-4) (d-2) f'(r)}{2 r}-\frac{1}{2} (d-2) f''(r)\nonumber
\\
& \mathbf{V}_{22}=\frac{4 (d-2) f(r)}{r^2}-\frac{(d-2) f'(r)}{r}+\frac{f'(r)^2}{f(r)}-f''(r)\nonumber
\\
& \mathbf{V}_{23}=\frac{(d-3) (d-2)}{r^2}-\frac{(d-2) (d+1) f(r)}{r^2}-\frac{(d-4) (d-2) f'(r)}{2
   r}-\frac{1}{2} (d-2) f''(r)\nonumber
\\
& \mathbf{V}_{33}=\frac{2 (d-4) (d-3) (d-2)}{r^2}-\frac{2 (d-5) (d-2)^2 f(r)}{r^2}-\frac{2 (d-2)^2 f'(r)}{r}\,.
\end{align}
Note in particular that neither $\mathbf{P}$ nor $\mathbf{V}$ are positive definite. This is to be expected, since we know that pure trace deformations (in the infinite cavity limit) will have a negative contribution to the Euclidean action.

In the infinite cavity limit, the trace modes and the traceless-tranverse modes decouple.  For this reason, we can then Wick rotate the trace modes freely. However, in the presence of a cavity they couple. Let us see this more explicitly.  We begin by separating the trace and trace-free parts of the metric by writing
\begin{subequations}
\begin{align}
&a=\frac{p}{d}-\hat{b}-(d-2)\hat{c}\,,
\\
&b=\hat{b}+\frac{p}{d}\,,
\\
&c=\hat{c}+\frac{p}{d}\,,
\end{align}
\end{subequations}
so that $h=p$.

However, in the canonical ensemble, perturbations must preserve the induced metric on the cavity walls (located at $r=r_0$). This amounts to imposing
\begin{equation}
a(r_0)=c(r_0)=0\, , 
\end{equation}
or 
\begin{equation}
\label{eq:CBCs}
p(r_0)=-d\,\hat c(r_0)\ {\rm and} \ p(r_0) = \frac{d}{d-1}\hat b(r_0). 
\end{equation}
Furthermore, at least with the symmetries imposed above, the part of this boundary condition that fixed the proper length of the Euclidean time circle at the value of $r$ where the sphere has radius $r_0$ is invariant under all diffeomorphisms, and is thus unaffected by the pure gauge modes\footnote{This is the key difference between finite $r_0$ and removing the cavity by taking $r_0 = \infty$.  As is well known, with asymptotically AdS boundary conditions the conformal rescaling of the boundary metric induced by any pure-trace mode can be compensated by a diffeomorphism.}.  Yet it is clear from \eqref{eq:CBCs} that the effect of any pure trace mode on the boundary conditions can be compensated by turning on a non-trace mode with appropriate $\hat c(r_0), \hat b(r_0)$ proportional to $p(r_0)$, and with real proportionality constants.  

As a result, when $p, \hat c, \hat b$ all take real values, there are allowed perturbations of $p$ with $p(r_0) \neq 0$.    But if we were to Wick rotate the contour of integration for $p$ so that $p(r_0)$ is imaginary while keeping the TT modes real, the (real) boundary conditions \eqref{eq:CBCs} would require $p(r_0)$ and the corresponding boundary values of the TT modes to vanish separately\footnote{While $\hat b$ and $\hat c$ can both be affected by pure-gauge modes, as described above there must be a gauge-invariant combination.  This combination is necessarily real due to the underlying reality of the physical system.}.  This would amount to imposing at least one extra boundary condition which, in particular, has no analogue in any real-time version of the system.  Such a condition would be highly suspect, and strongly motivates us to explore other proposals for defining the contour of integration such as that proposed in section \ref{sec:formalism}.

One can make the issue more concrete by choosing a gauge.  Let us consider in particular the de Donder gauge condition
\begin{equation}
\hat{\nabla}_a \bar{h}^{ab}=0\Rightarrow \hat{\nabla}_a h^{ab}-\frac{\hat{\nabla}^b h}{2}=0\,.
\end{equation}
This gives a first order differential equation that involves $\hat{b}$, $\hat{b}^\prime$, $\hat{c}$ and $\hat{p}^\prime$ which can be readily solved for $\hat{c}$:
\begin{equation}
\hat{c}=\frac{2}{2 f-r f^\prime} \left[\left(\frac{r f^\prime}{d-2}+f\right)\,\hat{b}+\frac{f\,r}{d-2}\hat{b}^\prime-\frac{f\,r}{2 d}p^\prime\right]\,.
\end{equation}
The boundary conditions then become
\begin{subequations}
\label{eq:DDBCs}
\begin{align}
& \hat{b}(r_0)=\frac{d-1}{d}p(r_0)
\\
&p'(r_0)-\frac{2 d}{d-2}\hat{b}'(r_0)-d\,\frac{p(r_0)}{r_0} \left[2+\frac{1}{d-2}\frac{r_0 f'(r_0)}{f(r_0)}\right]=0\, .
\end{align}
\end{subequations}
For $p,\hat b$ real, these define two boundary conditions as desired.  But for $p$ imaginary and $\hat b$ real, \eqref{eq:DDBCs} can be satisfied only if the real and imaginary parts vanish separately, effectively imposing {\it four} boundary conditions, which is two more than one would find in any Lorentz-signature analogue of the problem.

\subsection{A small interlude on numerical methods}
\label{sec:numerics}
A standard approach to minimizing the action Eq.~(\ref{eq:actionneg}) is to use an \emph{operator approach}. In such an approach, one does not work directly with the action, but instead studies the auxiliary problem
\begin{equation}
(\hat{\Delta}_L h)_{cd}=\lambda\,h_{cd},
\end{equation}
with $h$ chosen to satisfy the de Donder gauge. One then discretizes $\hat{\Delta}_L$ on a numerical grid and studies the resulting eigenvalues.

As advertised in section \ref{sec:further}, we will proceed differently here by  discretizing the action (\ref{eq:actionneg}) directly and then using the result to define a discretized \L operator.  Again, this allows us to easily incorporate complicated boundary conditions.  We will find this to be particularly useful when we study the microcanonical ensemble \cite{toappear}. We have checked that our action-based approach reproduces known results from the operator approach (including all that were reported in \cite{Headrick:2006ti}).

Numerically, it is useful to work with a compact coordinate $y\in[0,1]$ defined as
\begin{equation}
r=\frac{r_+}{1-\left(1-\frac{r_+}{r_0}\right)y^2}\,,
\label{eq:tra}
\end{equation}
so that the horizon is located at $y=0$ and the cavity walls at $y=1$. It is in these coordinates that we introduce a grid with $N+1$ discrete points. In this work, we will use spectral collocation methods on Gauss-Lobatto collocation points defined in terms of the coordinate $y$\footnote{We also solved this problem with second order finite differences, and the results are identical, though one has to use a larger number of points to recover the same accuracy, as expected.}. For the case at hand these are simply defined as
\begin{equation}
y_i = \frac{1}{2}\left[1+\cos\left(\frac{i \pi}{N}\right)\right]\,\quad\text{with}\quad i = 0,\ldots, N\,.
\end{equation}
Let $g(y)$ be a function, whose derivative we want to compute on the grid $\{y_i\}$. Let $\mathbb{Dg}$ be our approximation to the derivative of $g$ on the grid points $\{y_i\}$. It turns out that $\mathbb{Dg}$ can be computed via matrix multiplication $\mathbb{Dg}=\mathbb{D}\cdot \mathbb{g}$, with $\mathbb{g}_i \equiv g(y_i)$. Explicit expressions for the components of $\mathbb{D}$ can be found for instance in \cite{Dias:2015nua}.  Integration also is implemented by a similar linear operator, which can be taken to be given by the formulae in \cite{Canuto2010SpectralMF}.

Using this approach, one can write a discretization of the action in the form
\begin{equation}
\label{eq:actiondis}
\check{S}^{(2)}\approx \check{\mathbb{S}} \equiv \sum_{\tilde I,\tilde J=1}^{3(N+1)}\tilde{\mathbb{Q}}^{\tilde I} \tilde{\mathbb{M}}_{\tilde I \tilde J} \tilde{\mathbb{Q}}^{\tilde J},
\end{equation}
for some $\tilde {\mathbb{M}}_{\tilde I \tilde J}$, where the factor of $3$ in $3(N+1)$ is a result of the fact that we take  $\tilde{\mathbb{Q}}^{\tilde  I}=\{\mathbb{a},\mathbb{b},\mathbb{c}\}$ with $\mathbb{a}$, $\mathbb{b}$ and $\mathbb{c}$ being respectively discretized versions of $a$, $b$ and $c$.

As in section \ref{sec:further}, the tilde on
$\tilde{\mathbb{Q}}$ indicates that we have not yet imposed boundary conditions.  At the cavity walls we demand $a(1)=c(1)=0$, since we want the period of the time circle to remain unchanged, as well as the size of the $S^{d-2}$ sphere. At the horizon, we demand regularity, which in $y$ coordinates amounts to $a(0)=b(0)$, $a^\prime(0)=b^\prime(0)=c^\prime(0)=0$. It might appear that we are missing a boundary condition for $b$ at the cavity wall $y=1$. However, this may be resolved by imposing the de Donder gauge condition at this one point.  Doing so turns out to yield a Robin boundary condition
\begin{equation}
b^\prime(1)-a^\prime(1)-(d-2)c^\prime(1)+2\left(\frac{r_0}{r_+}-1\right) \left[2 (d-2)+\frac{r_0 f'\left(r_0\right)}{f\left(r_0\right)}\right] b(1)=0\,.
\end{equation}

We now discretize these boundary conditions using the scheme described above. For instance, the Neumann boundary condition for $a$ will appear as $\mathbb{D}_{N+1}\cdot \mathbb{a}=0$. We then use the boundary conditions to reduce Eq.~(\ref{eq:actiondis}) to a function of the degrees of freedom that remain unconstrained after imposing the boundary conditions. In particular, we set $\mathbb{a}_{N+1}=\mathbb{b}_{N+1}$, solve $\mathbb{D}_{N+1}\cdot \mathbb{a}=0$, $\mathbb{D}_{N+1}\cdot \mathbb{b}=0$ and $\mathbb{D}_{N+1}\cdot \mathbb{c}=0$ with respect to $\mathbb{a}_{N}$, $\mathbb{b}_{N+1}$ and $\mathbb{c}_{N+1}$, respectively. At the cavity, we solve with respect to $\mathbb{a}_1$, $\mathbb{b}_1$ and $\mathbb{c}_1$. Once these conditions are imposed, the sum in (\ref{eq:actiondis}) may be rewritten so that it sums only over the remaining unconstrained components $\mathbb{Q}$:
\begin{equation}
\check{\mathbb{S}} = \sum_{I,J=1}^{3 N-4} {\mathbb{Q}}^I\check{\mathbb{S}}_{,IJ}{\mathbb{Q}}^J\,.
\end{equation}
with ${\mathbb{Q}}=\{\mathbb{a}_2,\ldots,\mathbb{a}_{N-1},\mathbb{b}_2,\ldots,\mathbb{b}_{N},\mathbb{c}_2,\ldots,\mathbb{c}_{N}\}$. Because the boundary conditions are homogeneous linear relations, the action $\check{\mathbb{S}}$ remains a homogeneous quadratic function of the $\mathbb{Q}^I$.

As described in section \ref{sec:formalism}, we must now choose a metric $\mathbb{G}_{IJ}$ in order to use \eqref{eq:linop} define the linear operator $\mathbb{L}^I_J$.  As advertised in the introduction, in the continuum we choose the DeWitt$_{-1}$ metric for which
\begin{equation}
\lVert h \rVert^2 = (h,h) \equiv \frac{1}{32 \pi G}\int_{\mathcal{M}} \mathrm{d}^d x\sqrt{\hat{g}}\;h_{ab}\;\hat{\mathcal{G}}^{ab\;cd}\;h_{cd}\, ,
\label{eq:gnorm}
\end{equation}
where the factor of $(32 \pi G)^{-1}$ in front is chosen to agree with the conventions of \cite{Headrick:2006ti}. We now discretize \eqref{eq:gnorm} by following the same procedure used to discretize the action. The result is of the form 
\begin{equation}
\lVert h \rVert^2 \approx \sum_{I,J=1}^{3 N-4} {\mathbb{Q}}^I {\mathbb{G}}_{I J} {\mathbb{Q}}^J,
\end{equation}
where by construction both ${\mathbb{S}}_{,IJ}$ and ${\mathbb{G}}_{IJ}$ are symmetric in $I,J$.

Finally, we wish to solve for eigenvectors of the $\mathbb{L}^I_J$ defined by \eqref{eq:linop}.  However, we comment that this is equivalent to the so-called generalized eigenvalue problem 
\begin{equation}
\hat{\mathbb{L}}\cdot {\mathbb{Q}}_{\lambda}=\lambda\,\hat{\mathbb{G}}\cdot {\mathbb{Q}}_{\lambda}\, ,
\label{eq:sgep}
\end{equation}
where $\hat{\mathbb{L}}^I_J = \sum_K \delta^{IK} \mathbb{S}_{,IJ}$ and $\hat{\mathbb{G}}^I_J = \sum_K \delta^{IK} \mathbb{G}_{IJ}$ in terms of the Kronecker delta $\delta^{IK}$ and the metric $\mathbb{G}_{IJ}$.  Note in particular, since the operators $\hat{\mathbb{L}}, \hat{\mathbb{G}}$ are both self-adjoint with respect to the positive definite metric $\delta_{IJ}$.   The point of writing \eqref{eq:sgep} is that this formulation of our problem is found in many standard libraries of numerical methods.

\subsection{Results}
\label{sec:results}
\begin{figure}[h!]
    \centering
    \vspace{1cm}
    \includegraphics[width=0.85\textwidth]{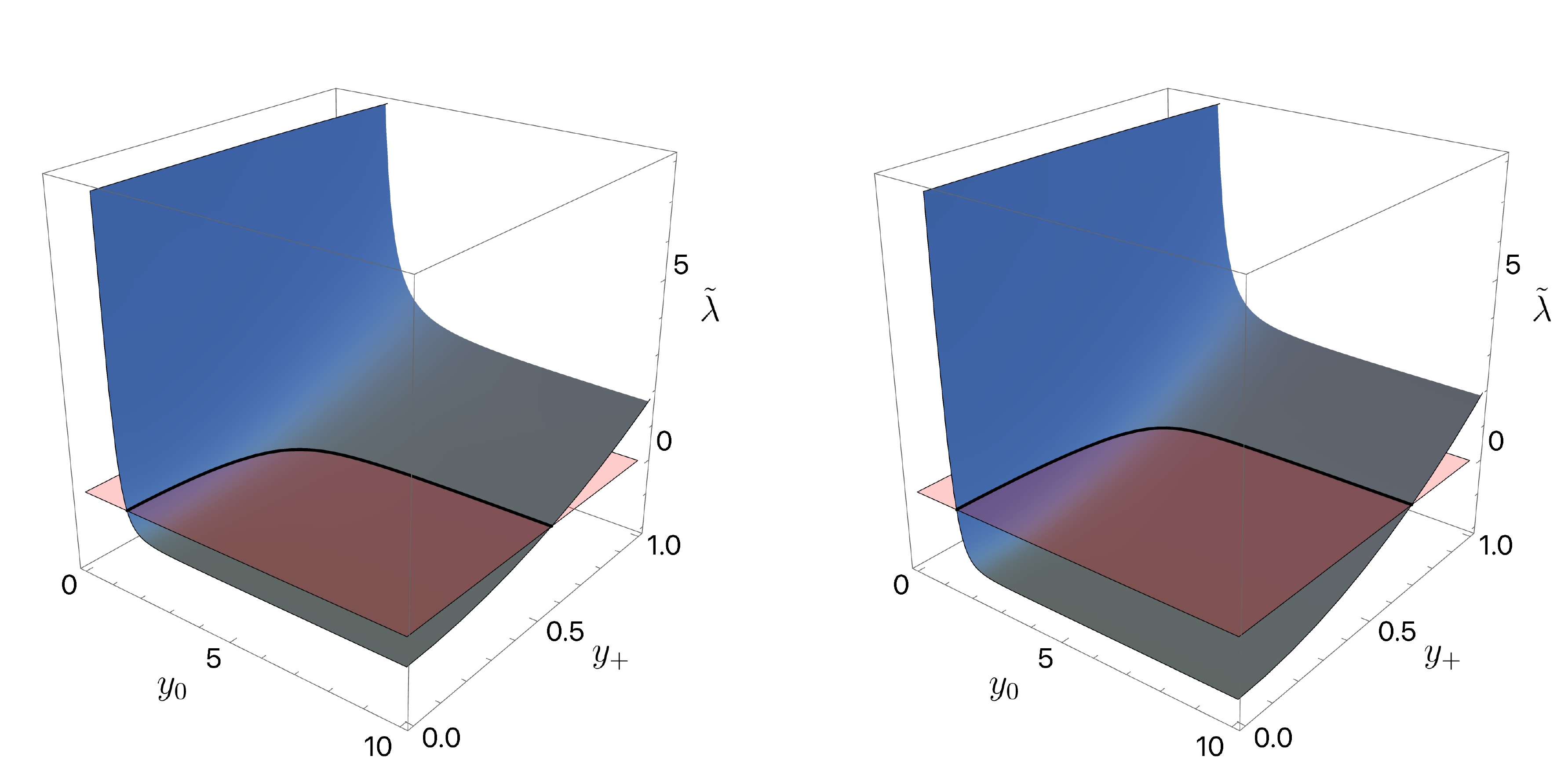}
    \caption{\label{fig:negative_mode_overall} The dimensionless lowest lying mode $\tilde{\lambda}\equiv \lambda\,r_+^2$ as a function of $y_0$ and $y_+$ for $d=4$ (left panel) and $d=5$ (right panel). The black line in each of the plots corresponds to the locus in moduli space where $\tilde{\lambda}=0$ and it coincides precisely with Eq.~(\ref{eq:crazyl}). To aid visualisation we also plot the plane $\tilde{\lambda}=0$ in red.}
\end{figure}

 Our findings are summarised in Fig.~(\ref{fig:negative_mode_overall}), where we  plot the dimensionless lowest lying $\tilde{\lambda}\equiv \lambda\,r_+^2$ as a function of $y_0\equiv r_0/r_+$ and $y_+\equiv r_+/\ell$ for $d=4$ (left panel) and $d=5$ (right panel).   In all cases, the mode with smallest ${\rm Re}\;\tilde{\lambda}$ has real $\tilde{ \lambda}$. We also plot the plane $\tilde{\lambda}=0$ in red, so that it is apparent when the mode becomes positive. The solid black curve in each of the plots is given by Eq.~(\ref{eq:crazyl}) and it precisely matches the locus where $\tilde{\lambda}$ changes sign. In the limit where $y_0$ becomes large we recover the results of \cite{Gross:1982cv,Prestidge:1999uq,Hubeny:2002xn,Asnin:2007rw,Dias:2010eu,Dias:2015pda}. To our knowledge this work is the first to report the lowest lying mode in AdS at finite $y_0$. 

We now come to the issue of the norm of the mode under $\hat{\mathcal{G}}$. Any metric perturbation can be decomposed as a sum of a traceless component $\tilde{h}_{ab}$ and a pure trace part $\phi$:
\begin{equation}
h_{ab}=\tilde{h}_{ab}+\frac{1}{d}\hat{g}_{ab}\,\phi\,.
\end{equation}
The metric $\hat{\mathcal{G}}$ is such that $\tilde{h}_{ab}$ and $\hat{g}_{ab}\,h$ are orthogonal to each other. That is to say
\begin{equation}
\tilde{h}_{ab} \hat{G}^{ab\,cd}\hat{g}_{cd}=0\,.
\end{equation}
This, in turn, implies that the norm defined in (\ref{eq:norm}) may be written in the form
\begin{subequations}
\begin{equation}
\lVert h \rVert^2 = \frac{1}{32 \pi G}\left(\int_{\mathcal{M}} \mathrm{d}^d x\sqrt{\hat{g}}\;\tilde{h}_{ab}\;\tilde{h}^{ab}- \int_{\mathcal{M}} \mathrm{d}^d x\sqrt{\hat{g}}\;\phi^2\right)\equiv\lVert \tilde{h} \rVert^2_{\infty}-\lVert \phi \rVert^2_{\infty}\,,
\end{equation}
where we have defined
\begin{equation}
\lVert \tilde{h} \rVert^2_{\infty}\equiv \frac{1}{32 \pi G}\int_{\mathcal{M}} \mathrm{d}^d x\sqrt{\hat{g}}\;\tilde{h}_{ab}\;\tilde{h}^{ab}>0
\end{equation}
and
\begin{equation}
\lVert \phi \rVert^2_{\infty}\equiv \frac{1}{32 \pi G}\frac{d-2}{2d}\int_{\mathcal{M}} \mathrm{d}^d x\sqrt{\hat{g}}\;\phi^2>0\,.
\end{equation}
\end{subequations}
It thus follows that positivity of $\lVert h \rVert^2$ is equivalent to positivity of 
\begin{equation}
\eta\equiv 1-\frac{\lVert \phi \rVert^2_{\infty}}{\lVert \tilde{h} \rVert^2_{\infty}}\, .
\label{eq:eta}
\end{equation}

\begin{figure}[h!]
    \centering
    \vspace{1cm}
    \includegraphics[width=0.85\textwidth]{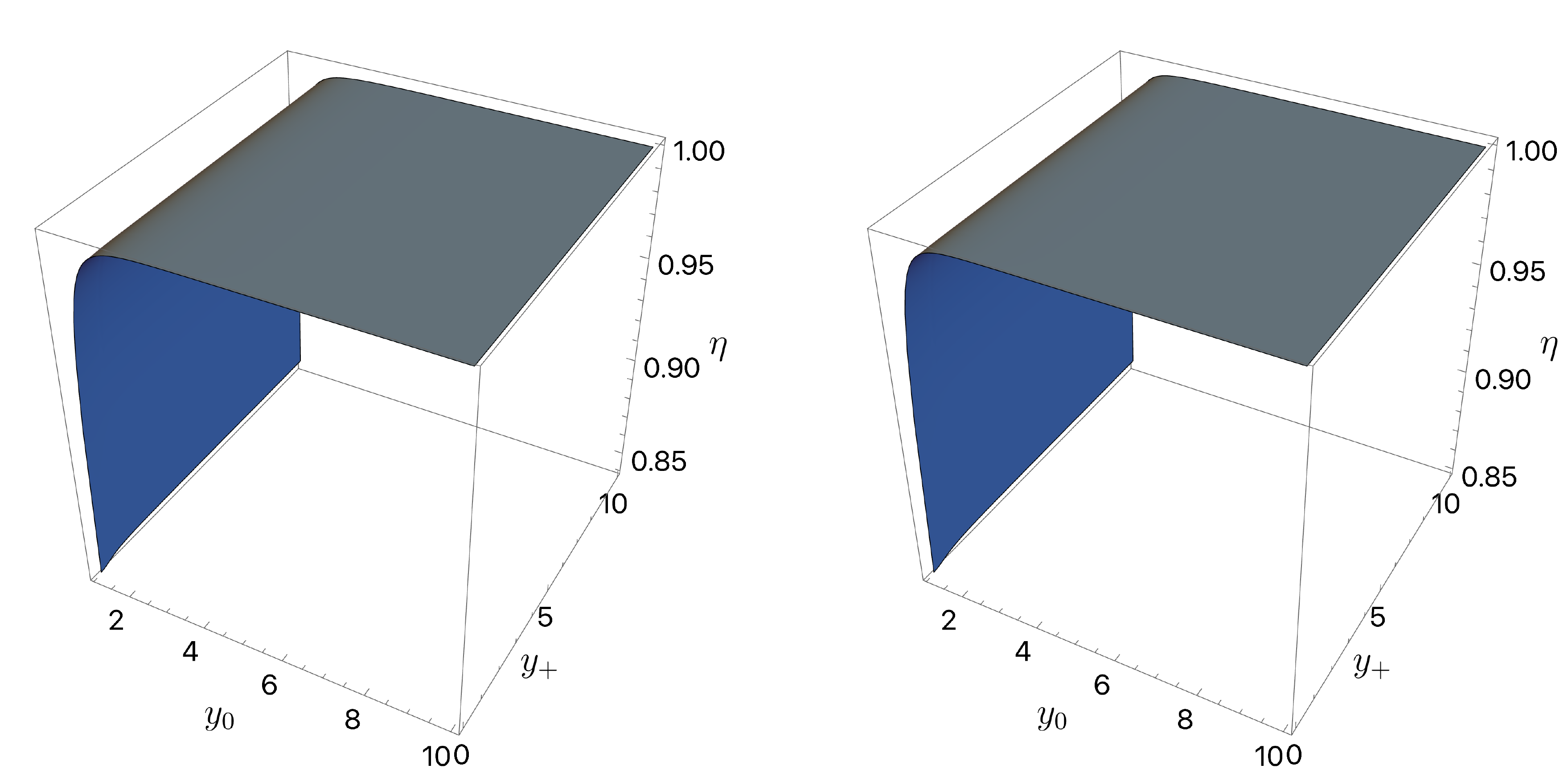}
    \caption{\label{fig:negative_norm_overall} $\eta$, defined in (\ref{eq:eta}), as a function of $y_0$ and $y_+$ for $d=4$ (left panel) and $d=5$ (right panel). From the positivity of $\eta$ we conclude that the norm of the lowest lying mode is positive definite, and as such no Wick rotation is necessary.}
\end{figure}

In Fig.~(\ref{fig:negative_norm_overall}) we plot $\eta$ as a function of $y_0$ and $y_+$ for $d=4$ (left panel) and $d=5$ (right panel). We see that the norm is positive definite everywhere, so that, according to our rule, we should \emph{not} Wick rotate. $\eta$ also reveals another expected result, namely, the fact that the trace mode decouples from the traceless-transverse part of the metric. This is seen in Fig.~(\ref{fig:negative_norm_overall}), where we see $\eta$ approaching unity when the cavity is removed, \emph{i.e.} $y_0\to+\infty$.

Having established that the Schwarzschild AdS black hole is \emph{unstable} whenever the lowest lying mode is negative, we turn turn to the issue of the details of the spectrum.

\begin{figure}[h!]
    \centering
    \vspace{1cm}
    \includegraphics[width=0.85\textwidth]{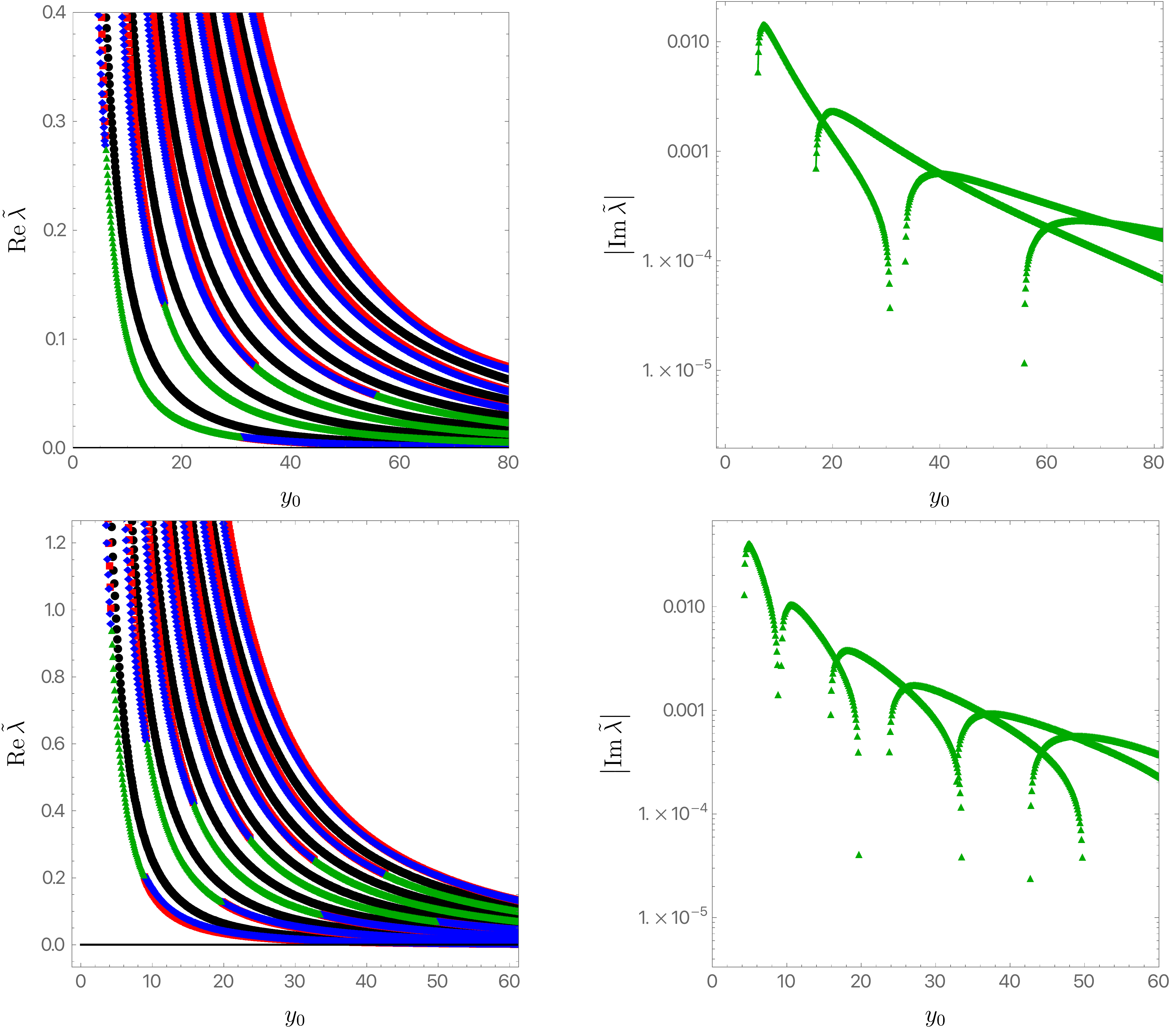}
    \caption{\label{fig:overtones4D} The real part (left column) and the absolute value of the imaginary part (right column) of the excited modes as a function of $y_0$ and $y_+=0$. The top row has $d=4$, while the bottom row has $d=5$. The colour coding is as follows: green triangles are non-gauge modes with complex eigenvalues; blue diamonds are non-gauge modes with negative norm under $\hat{\mathcal{G}}$; red squares are non-gauge modes with positive norm under $\hat{\mathcal{G}}$ and the black disks are pure gauge modes. The plot on the right panel is presented in a logarithmic scale, and each green triangle has a two-fold degeneracy since complex modes come in conjugate pairs.}
\end{figure}

For simplicity, we start by studying the case with $y_+=0$, \emph{i.e.} the case with a vanishing cosmological constant, and we restrict to excited modes (\emph{i.e.}, we leave aside the lowest-lying mode that we have already studied). In Fig.~\ref{fig:overtones4D} we plot the eigenvalues corresponding to the first twenty excited modes as a function of $y_0$. The colour coding is as follows: green triangles are non-gauge modes with complex eigenvalues; blue diamonds are non-gauge modes with negative norm under $\hat{\mathcal{G}}$; red squares are non-gauge modes with positive norm under $\hat{\mathcal{G}}$ and the black disks are pure gauge modes. Modes are classified as gauge vs. non-gauge by comparing the modified action $\check{S}{}^{(2)}$ to the original action ${S}{}^{(2)}$, with vanishing of the latter (to numerical precision) indicating a gauge-mode while for non-gauge modes the values agree.  As a consistency check, we also verify that the non-gauge modes are precisely those that satisfy the De Donder gauge condition as expected.

In the left column of Fig.~\ref{fig:overtones4D} we plot the real part of the eigenvalues, and on the right column we plot the absolute value of the imaginary part of the complex eigenvalues using a logarithmic scale. The top row has $d=4$, while the bottom row has $d=5$.

The first and most important thing we note is that all excited modes have $\mathrm{Re}\,\tilde{\lambda}>0$, thus establishing stability with respect to all excited modes for all $y_0,y_+$. Again, we remind the reader that this sets aside the lowest-lying mode studied earlier. 

The second observation is simply that complex modes exist, in the sense that some eigenvalues have non-zero imaginary parts. In fact, because they must come in complex conjugate pairs, the spectrum exhibits \emph{bubbles} at the edges of which non-gauge modes with positive and negative norms merge together to form a complex mode (which, as described in section \ref{sec:formalism}, always have zero norm). Each green triangle (which denotes data associated with a complex eigenvalue) has a two-fold degeneracy because complex mode appear in conjugate pairs.

We now further address the issue of completeness of the spectrum. We first note that in generic regions of parameter space our numerics yield exactly $3N-4$ distinct eigenvalues (here we are counting $\lambda$ and $\lambda^*$ as distinct eigenvalues when $\lambda$ is complex). This is precisely the dimension of the space of all independent unconstrained perturbations $\mathbb{Q}$.  As explained in section \ref{sec:Wick}, this is sufficient to show that we can construct $3N-4$ orthogonal eigenvectors, thus showing that $\mathbb{L}$ is digonalizable and that the eigenvectors of $\mathbb{L}$ span the full space of perturbations in generic regions of parameter space.

However, one might wonder what happens at the special values of $y_0, y_+$ where the eigenvalues become degenerate, \emph{i.e.} where a bubbles first forms or where it disappears. For simplicity, we will restrict detailed discussion to the case with vanishing cosmological constant, but we find similar results when $\Lambda\neq0$. Since the pure gauge modes never become complex, they are absent from this discussion. Let us denote by region $I$, the range of $y_0$ where modes are all real. In this region we have negative-norm modes and positive-norm modes. Let denote by region $II$ the range of $y_0$ where complex modes exist. The question is then what happens to these modes as we cross from region $I$ into region $II$ and vice-versa. We find that, with our normalisation for the complex modes, the limit of the negative-norm eiegenmode from region $I$ agrees to numerical precision with the limit of the imaginary part of the eigenmode from region $II$, while the limit of the positive-norm eigenmode in region $I$ agrees with the limit of the real part of the complex eigenmode from region $II$.  Furthermore, the limits of the positive- and negative-norm eigenmodes from region $I$ are linearly independent.   This means that, as previously advertised, the eigenmodes continue to span the space of all modes even when the eigenvalues become degenerate at the edges of the bubbles.   Furthermore, the continuity between these modes in passing from region $I$ to region $II$ ensures that our Wick rotation is well defined throughout the entire parameter space, including the critical points when the spectrum becomes degenerate.

\begin{figure}[h!]
    \centering
    \vspace{1cm}
    \includegraphics[width=0.9\textwidth]{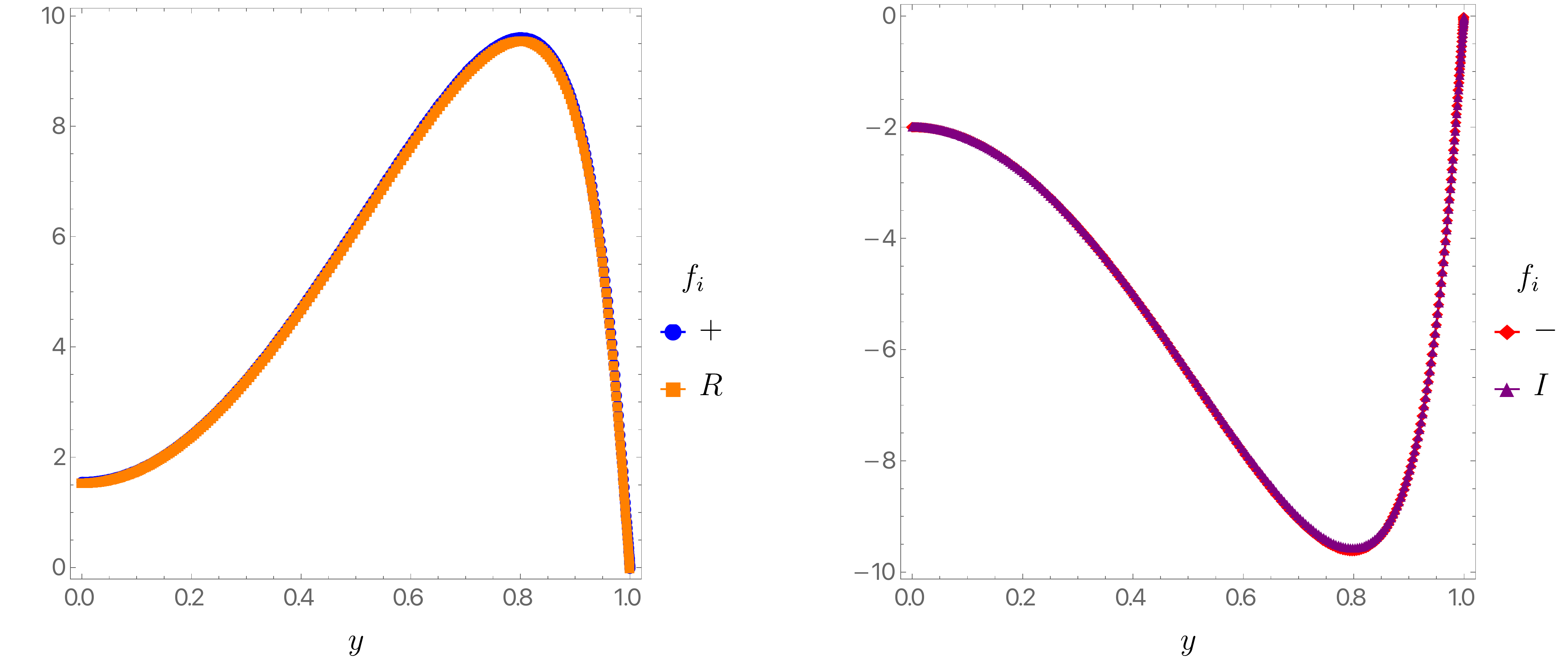}
    \caption{\label{fig:comp} {\bf Left panel}: Two eigenfunctions are plotted as a function of $y$. The blue disks represent $f_+$, defined as the limit of the positive-norm eigenfunction from region $I$, and the orange squares represent $f_R$, defined as the limit of the real part of the complex eigenfunction (using our normalisation) from region $II$. {\bf Right panel}: Two eigenfunctions are plotted as a function of $y$. The red diamonds represent $f_-$, defined as the limit of the negative-norm eigenfunction from region $I$, and the purple triangles represent $f_I$, defined as the limit of the imaginary part of the complex eigenfunction (using our normalisation) from region $II$. Region $I$ is defined as $y_0\lesssim y_{\mathrm{b}}=6.0285$, while region $II$ is defined as $y_0\gtrsim y_{\mathrm{b}}$.}
\end{figure}

As a typical example, let us investigate the region of moduli space near $y_0\approx y_{\mathrm{b}}=6.0285$ with $y_+=0$. This corresponds to the region in the top row, right column, of Fig.~3 where the first bubble on the left forms. We use $f_{\pm}$ to denote the positive-norm and negative-norm eigenfunctions, respectively, associated with the perturbed metric function $a$ for $y_0\lesssim y_{\mathrm{b}}$. Additionally, let us define $f_{R}$ ($f_I$) to be the real (imaginary) part of the complex eigenfunction associated with the perturbed metric function $a$ for $y_0\gtrsim y_{\mathrm{b}}$. In Fig.~\ref{fig:comp} we plot on the left panel $f_+$ (blue disks) and $f_R$ (orange squares), while on the right panel we plot $f_-$ (red diamonds) and $f_I$ (purple triangles). We see that
\begin{equation}
\lim_{y_0\to y_{\mathrm{b}}^-} f_+=\lim_{y_0\to y_{\mathrm{b}}^+}f_R\quad \text{and}\quad \lim_{y_0\to y_{\mathrm{b}}^-} f_-=\lim_{y_0\to y_{\mathrm{b}}^+}f_I
\end{equation}
as claimed in the preceding paragraph.

Let us now return to describing general features of the spectrum, this time with a non-zero cosmological constant.  Perhaps surprisingly, we find that complex eigenvalues exist for all values of $y_+$, and in particular even if the black hole is very large compared to the AdS scale. However, as we describe below, at large $y_+$ a given excitation becomes complex only in a very narrow window of $y_0$. 

\begin{figure}[h!]
    \centering
    \vspace{1cm}
    \includegraphics[width=0.9\textwidth]{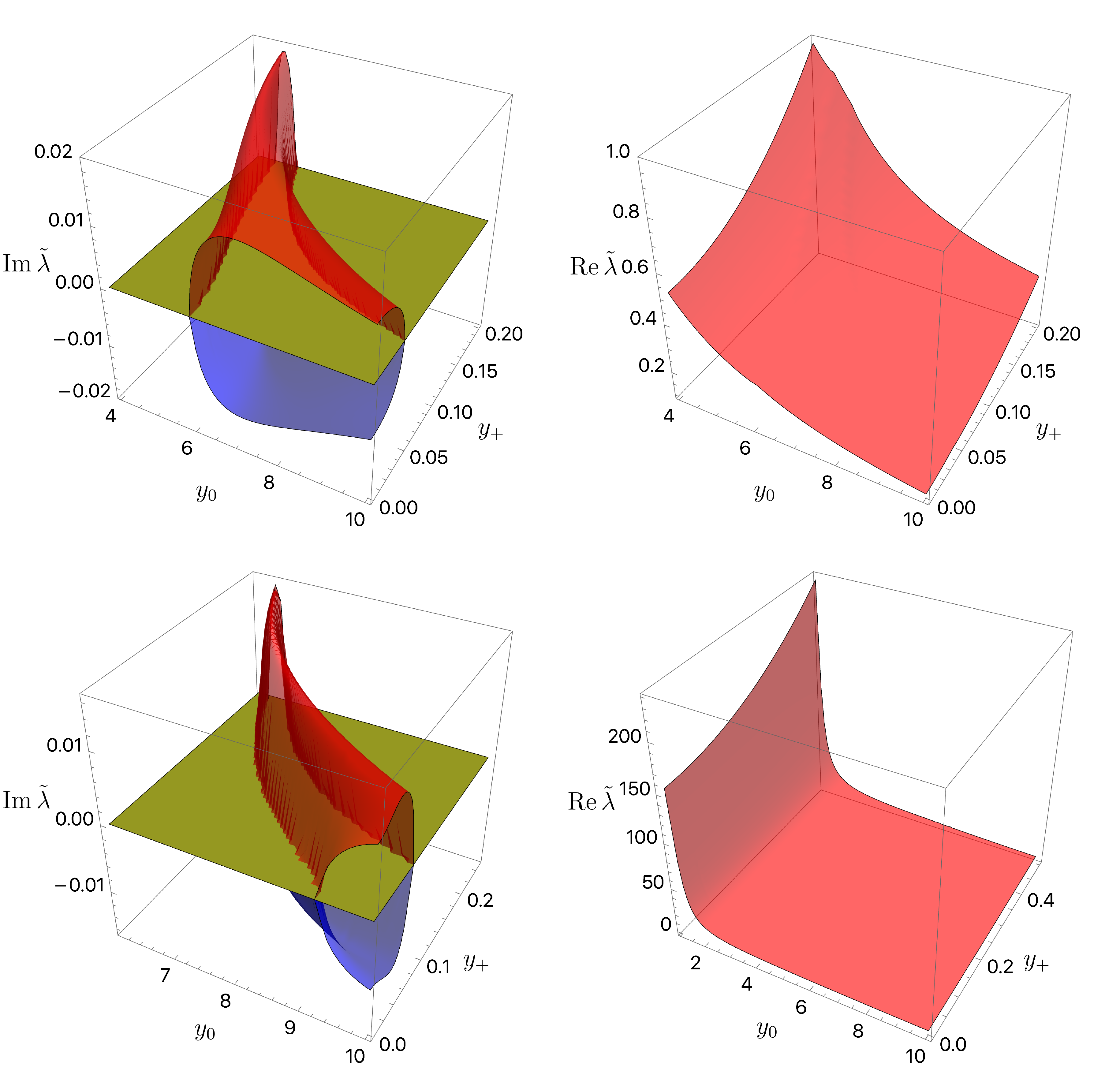}
    \caption{\label{fig:complex}The imaginary part (left column) and the imaginary part (right column) of the first excited mode in a region of $y_0$ and $y_+$ where the mode is complex. The top row has $d=4$, while the bottom row has $d=5$. The green horizontal plane on the left column sits at $\tilde{\lambda}=0$ and is just there for illustration purposes. The mode with $\mathrm{Im}\,\tilde{\lambda}>0$ is represented in red, while for $\mathrm{Im}\,\tilde{\lambda}<0$ we represent the mode in blue.}
\end{figure}

It is computationally challenging to monitor what happens to \emph{all} the bubbles we found with $y_+=0$ as we increase $y_+$, so we instead focus on the bubble in $d=4$ and $d=5$ that starts at the smallest value of $y_0$ and follow its dependence on $y_+$. This is the bubble associated with the first pair of positive/negative norm excitations.  In Fig.~\ref{fig:complex} we plot the imaginary part (left column) and real part (right column) of the eigenvalues for the first excited mode in the region where $\tilde{\lambda}$ is complex. The top row has $d=4$, while the bottom row has $d=5$. The green horizontal plane on the left column lies at $\tilde{\lambda}=0$ and is included for illustration purposes only. The mode with $\mathrm{Im}\,\tilde{\lambda}>0$ is represented in red, while for $\mathrm{Im}\,\tilde{\lambda}<0$ we represent the mode in blue. From this figure it is clear that the width (measured in terms of $y_0$) of the bubble becomes shrinks as we increase $y_+$.

\begin{figure}[h!]
    \centering
    \vspace{1cm}
    \includegraphics[width=0.9\textwidth]{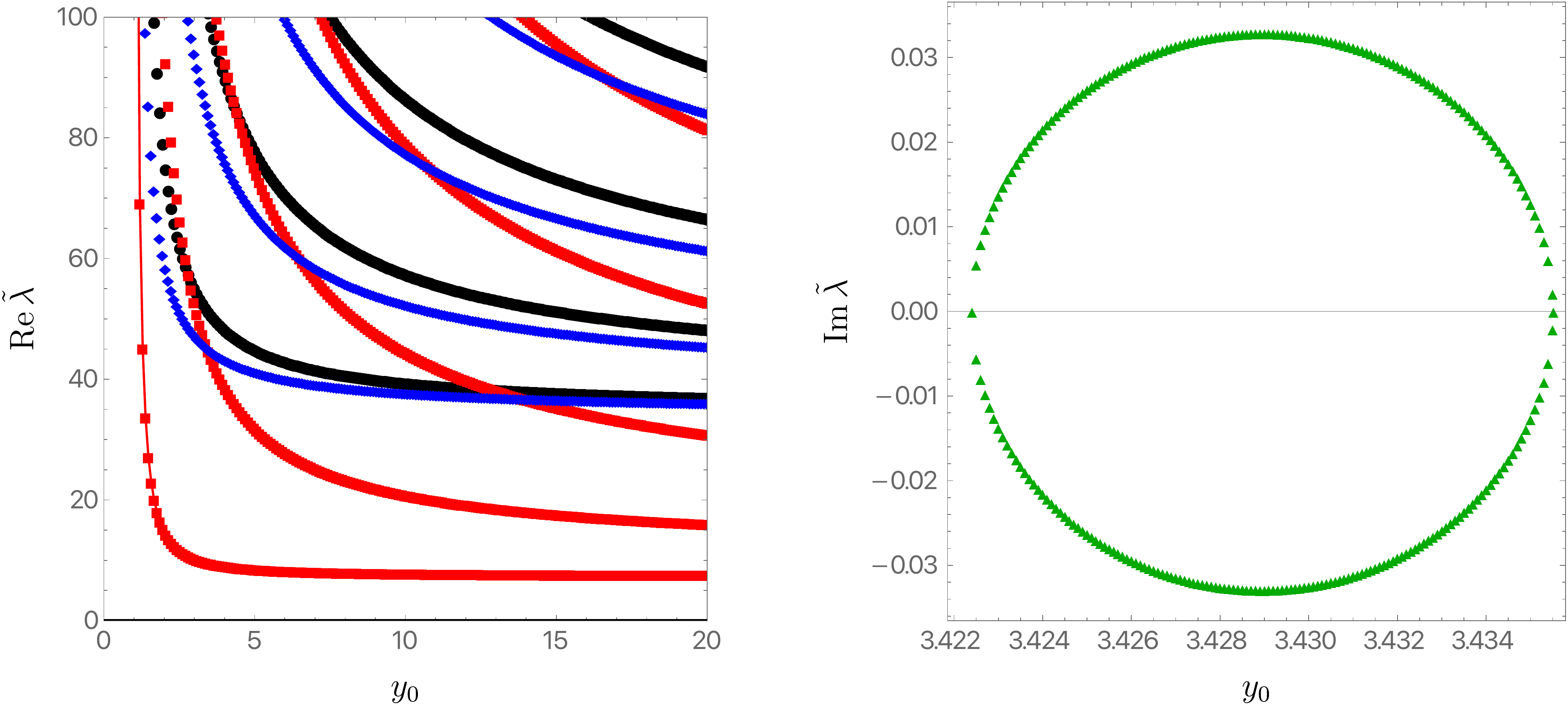}
    \caption{\label{fig:small}The real part (left panel) of the excited modes as a function of $y_0\in(1.1,20)$ and $y_+=20$. In the right panel we plot the imaginary part of $\tilde{\lambda}$ in a magnified description of the region where the first excited non-gauge mode with negative norm crosses the first excited non-gauge mode with positive norm. The colour coding is as follows: green triangles are non-gauge modes with complex eigenvalues; blue diamonds are non-gauge modes with negative norm under $\hat{\mathcal{G}}$; red squares are non-gauge modes with positive norm under $\hat{\mathcal{G}}$ and the black disks are pure gauge modes.}
\end{figure}

From the behaviour as we increase $y_+$, it might appear that the bubble will disappear altogether at some threshold value of $y_+$. However, we find that this is \emph{not} the case. Instead, it appears that for any finite value of $y_+$ we can find a small bubbles where the given modes becomes complex, though the width of the bubbles becomes incredibly small. In Fig.~\ref{fig:small} we plot the real part of $\tilde{\lambda}$ (left panel) for $d=4$, as a function of $y_0\in(1.1,20)$ and $y_+=2$ using the same colour coding as in Fig.~\ref{fig:overtones4D}. Examining the left panel it appears that complex modes do not exist. However, if we zoom in on the region where positive and negative modes appear to cross we see that complex bubbles do exist (right panel) in an incredibly narrow region of $y_0$. We have confirmed this picture for values of $y_+$ as large as $100$.

It would be interesting to explore these structures further in the future.  One may expect that more complicated black holes, and in particular those associated with additional free parameters, lead to even more intricate structures.
\section{Discussion \& Conclusions}
\label{sec:disc}

The main point of our work above was to generalize the proposal of Gibbons, Hawking, and Perry for the contour of integration in the path integral describing fluctuations about a saddle in Euclidean quantum gravity.  This was motivated by consideration of Euclidean gravity in a reflecting cavity, where boundary conditions couple the pure-trace and transverse-traceless (TT) modes, preventing us from Wick-rotating one the former without also Wick-rotating the latter.    However, by making use of the DeWitt$_{-1}$ metric,  the quadratic action can still be said to define a linear operator $\mathbb{L}$.  When this operator can be diagonalized, its eigenmodes
can be used to specify an integration contour that reduces to the Gibbons-Hawking-Perry proposal in their context.  The result is an eigenmode is stable precisely when its eigenvalue $\lambda$ satisfies ${\rm Re}\,\lambda > 0.$
We show numerically that $\mathbb{L}$ can be diagonalized for cavity fluctuations about Euclidean Schwarzschild-AdS black holes, and that our recipe reproduces the stability/instability of large/small black holes expected from the black hole thermodynamics studied in appendix \ref{sec:AdStherm}.    While the proposal thus satisfies an important physical consistency requirement, it remains somewhat \emph{ad hoc}.  For that reason we describe the proposal as a `rule of thumb.'

Because our $\mathbb{L}$ is just the \L operator (or a discretization thereof), the stability condition ${\rm Re}\,\lambda \ge 0$ is precisely the same as that proposed in \cite{Headrick:2006ti} based on Ricci flow (or a generalization thereof to include a cosmological constant).  Indeed, while \cite{Headrick:2006ti} did not discuss the possibility of complex eigenvalues, when they exist the condition for stability of their flow is again ${\rm Re}\,\lambda \ge 0$.

There are many future directions to explore. 
In particular, the conjectured generality of our rule of thumb opens up a wide arena of related contexts for the community to investigate.  In addition to adding charge, and perhaps rotation, one is also free to consider all manner of cavities (e.g., Schwarzschild black holes in rectangular or ellipsoidal boxes), or perhaps infinite-volume AdS systems with complicated boundary conditions. It would be interesting to check that $L$ remains diagonalizable in all cases, and that our rule of thumb continues to reproduce expectations from black hole thermodynamics\footnote{In a similar yet different direction, forthcoming work \cite{toappear} will use the rule of thumb described here to study saddle points for Euclidean path integrals describing the microcanonical ensemble for gravitational systems.}. It also remains to check our assumption that angular momentum does indeed increase $\rm{Re} \ \lambda$ and thus that the vector- and tensor-derived modes are indeed stable, to investigate whether such higher angular momentum modes become complex, and to similarly study modes that break time-translation symmetry.  Finally, one would also like to have a more general abstract argument for such agreement along the lines of section section 4.3 of \cite{Dias:2010eu}, but somehow showing that the negative mode discussed there is not affected by the Wick rotation defined by our rule of thumb.

Perhaps more interesting is the question of whether choosing a different value of the DeWitt parameter $\lambda_{DW} \neq -1$ continues to give physically viable results.  A priori it would appear that our procedure is well-defined for any $\lambda_{DW}$, but since $\lambda_{DW}$ affects the definition of $L$ it is natural to expect that the stability/instability of large/small black holes will be reproduced for at most one value of $\lambda_{DW}$.  However, this remains to be checked, and it is worth noting that in the absence of a cavity wall the decoupling of trace and TT perturabtions means that our rule of thumb would yield identical results for all $\lambda_{DW}$.

Finally, while the results of our Euclidean analysis seem quite satisfactory, it remains to find a first-principles derivation of our rule of thumb.   We expect the Lorentz-signature path integral to provide a useful starting point for such an analysis since, as an oscillatory integral, it should be well-defined in a distributional sense without any Wick rotation.  Allowed deformations of the contour of integration for this path integral might then be used to define the correct Euclidean prescription. We hope to at least partially address this issue in future work.

\paragraph{Acknowledgments}
We would like to thank \'Oscar~Dias, Matthew~Headrick and Toby~Wiseman for providing useful comments on an earlier version of this manuscript. D.~M. was supported by NSF grant PHY-2107939  and by funds from the University of California. J.~E.~S has been partially supported by STFC consolidated grant ST/T000694/1.

\appendix

\section{Properties of the de Donder Gauge}
\label{sec:DDgauge}

The gauge transformations of linearized gravity may be written in the form
\begin{equation}
h_{ab} \rightarrow h_{ab} + \nabla_a \xi_b + \nabla_b \xi_a
\end{equation}
for any one-form $\xi_a$.  As usual we define gauge transformations to be such that the tangential projection of $\xi^a$ into the cavity wall vanishes. In our case, we also fix the cavity wall to be at the coordinate location $r=r_0$.  The remaining gauge transformations then have $\xi^r(r_0)=0$, so that $\xi^a$ vanishes entirely at the cavity wall.

  As a result, we refer to perturbations of the form
$h^{gauge}_{ab} = \nabla_a \xi_b + \nabla_b \xi_a$ as pure gauge modes whenever $\xi^a|_{r=r_0}=0$.

As in the main text, we restrict attention to transformations that preserve explicit spherical symmetry and time-translation symmetry, so that $\xi^a$ has only $r$ and $\tau$ components and both are independent of $\tau$. 
We denote the space of such pure gauge modes by $V$.

It will be useful to understand the space $V^\perp$ of perturbations that are orthogonal to $V$ in the DeWitt$_{-1}$ inner product.  Using \eqref{eq:DWIP} and \eqref{eq:DW-1}, the statement $h_{ab} \in V^\perp$ takes the form

\begin{eqnarray}
0 &=& 32\pi G (h,g^{gauge} ) = \int_{\mathcal{M}}\mathrm{d}^d x\,\sqrt{\hat{g}}\, \left( 2 h_{ab} \nabla^a \xi^b   - h \nabla_a \xi^a\right)\nonumber \\
&=&  -2 \int_{\mathcal{M}}\mathrm{d}^d x\,\sqrt{\hat{g}}\, \left( \nabla_a h_{ab}- \frac{1}{2} \nabla_b h \right) \xi^b,
\label{eq:Vperp}
\end{eqnarray}
where we used the condition $\xi^a=0$ to drop a potential boundary term. 
It follows that $V^\perp$ contains precisely those modes that satisfy our boundary conditions and the de Donder condition
\begin{equation}
\label{eq:DDC}
\nabla^a h_{ab}- \frac{1}{2} \nabla_b h =0.
\end{equation}

Let us consider in particular the $\tau$ component of \eqref{eq:DDC}, which reads  
\begin{equation}
\partial_r h_{r\tau} + \frac{d-2}{r} h_{r\tau} + (\partial_r \ln f) h_{r\tau}=0\,..
\end{equation}
This is an ordinary differential equation for $h_{r\tau}$ which we can readily solve
\begin{equation}
h_{r\tau}=\frac{C_1}{f r^{d-2}}\,,
\end{equation}
where $C_1$ is a constant of integration. Regularity at the horizon, where $f$ vanishes, thus requires $C_1=0$ and thus $h_{r\tau}=0$ for all $r$. This part of the de Donder condition is thus simple to impose analytically.

Finally, we would like to show that every perturbation can be written as the sum of a mode in $V$ and a mode in $V^\perp$, and that in fact this decomposition is unique.  The fact that the DeWitt$_{-1}$ metric is not positive definite means that we cannot simply take this for granted, as complications arise if the induced metric on $V$ is degenerate.  To show that this is not the case, note that a degenerate induced metric on $V$ would require the existence of a $\tilde \xi^a$ for which the pure-gauge mode $h_{ab} = \nabla_a \tilde{\xi}_b + \nabla_b \tilde{\xi}_a$ is orthogonal to all modes in $V$; \emph{i.e.}, for all $\xi_a$, f using $h_{ab} = \nabla_a \tilde{\xi}_b + \nabla_b \tilde{\xi}_a$ would satisfy \eqref{eq:Vperp}.  Thus $h_{ab} = \nabla_a \tilde{\xi}_b + \nabla_b \tilde{\xi}_a$ satisfies the de Donder condition which yields 
\begin{equation}
\label{eq:nablaxi}
\nabla^2 \tilde \xi_a + \frac{2\Lambda}{d-2}\tilde \xi_a=0.
\end{equation}
But since $\tilde  \xi^a$ vanishes at the cavity wall, \eqref{eq:nablaxi} then requires
\begin{align}
\label{eq:resid}
0 &= \int_{\mathcal{M}} \mathrm{d}^dx\,\sqrt{\hat{g}} \tilde  \xi^a\,\left[\nabla^2 \tilde \xi_a+\frac{2\Lambda}{d-2}\tilde \xi_a\right]  \nonumber
\\
& = - \int_{\mathcal{M}}  \mathrm{d}^dx\,\sqrt{\hat{g}}\,\left[(\nabla_b \tilde \xi^a)( \nabla^b \tilde \xi_a)-\frac{2\Lambda}{d-2}\tilde \xi_a \tilde \xi^a\right].
\end{align}
For a Riemannian metric and for $\Lambda\leq0$, the right-hand side is positive definite. If $\Lambda<0$, the second term demands $\tilde \xi^a \tilde \xi_a=0$, and thus $\tilde \xi=0$ for a Riemannian metric. If $\Lambda=0$,
the equality can hold only if $\nabla_b \tilde \xi_a=0$; \emph{i.e.}, if $\tilde \xi_b$ is covariantly constant. But this would imply $h_{ab}=0$, so the supposed perturbation is trivial.  It follows that the DeWitt$_{-1}$ metric induces a non-degenerate metric on $V$, and that $V$ and $V^\perp$ intersect only on the zero perturbation.  It also follows that any allowed perturbation can be written uniquely as the sum of a pure-gauge mode in $V$ and a perturbation in $V^\perp$.  In particular, the projection into $V^\perp$ defines a good gauge fixing of our perturbations.

\section{Thermodynamics of Schwarzschild-AdS black holes in a spherical box}
\label{sec:AdStherm}

We begin with a brief review in section \ref{sec:nocav} of the  thermodynamics of Schwarzschild-AdS black holes in infinite volume.  The corresponding system inside a spherical box is then analyzed in section \ref{sec:addcav}.

\subsection{Review of the infinite volume case}
\label{sec:nocav}

We study the meric \eqref{eq:SAdSmetric} with the conventions stated at the beginning of section \ref{sec:SAdS} and with a fixed boundary metric at the AdS conformal boundary.  
Recall that for small enough real and positve values of $\beta_*$ there are two corresponding values of $r_+$,
\begin{equation}
r_+^{(\pm)}(\beta) =\frac{\ell^2}{d-1} \left[\frac{2 \pi }{\beta }\pm\sqrt{\left(\frac{2 \pi }{\beta }\right)^2-\frac{(d-3) (d-1)}{\ell^2}}\right],
\label{eq:sl}
\end{equation}
associated respectively with the large and small Euclidean Schwarzschild-AdS black holes.  There is a maximum value of $\beta_*$, which can be found by equating the argument of the square root in Eq.~(\ref{eq:sl}) to zero. This is the point at which the large and small branches of the Euclidean Schwarzschild-AdS black hole solution coincide.

The entropy of the Euclidean Schwarzschild-AdS black hole is given by
\begin{subequations}
\begin{equation}
S = \frac{\Omega_{d-2}\,r_+^{d-2}}{4 G_d}\,,
\label{eq:entropyschwarzschildAdS}
\end{equation}
where $\Omega_n$ is the area of a unit radius $n-$sphere, \emph{i.e.}
\begin{equation}
\Omega_n = \frac{2 \pi^{\frac{n+1}{2}}}{\Gamma\left(\frac{n+1}{2}\right)}\,,
\end{equation}
where $\Gamma(z)$ is the Gamma function. From the entropy (\ref{eq:entropyschwarzschildAdS}) it is a straightforward exercise to compute the specific heat $C$
\begin{equation}
C=-\beta \frac{\partial S}{\partial \beta}=\frac{(d-2) \pi  \Omega _{d-2}}{(d-1) r_+^2-(d-3) \ell^2}\,\frac{r_+^{d-1} \ell^2}{G_d}\,.
\end{equation}
\end{subequations}%
Local thermodynamic stability in the canonical ensemble then demands $C>0$, which translates to
\begin{equation}
r_+\geq r^{\star}_+\equiv  \sqrt{\frac{d-3}{d-1}}\,\ell\Rightarrow \beta_*\leq\beta^{\star}\equiv \frac{2 \pi }{\sqrt{d-3} \sqrt{d-1}}\ell\,.
\label{eq:trans}
\end{equation}
We also record the fact that
the energy of the Euclidean Schwarzschild-AdS solution\footnote{Here we ignore the Casimir energy present in odd spacetime dimensions.} is given by \cite{Balasubramanian:1999re,deHaro:2000vlm,Gibbons:2004ai}
\begin{equation}
E = \frac{(d-2) \Omega _{d-2} r_+^{d-3}}{16 \pi  G_d} \left(\frac{r_+^2}{\ell^2}+1\right)\,,
\end{equation}
which can be shown to satisfy the first law of black hole mechanics
\begin{equation}
\beta_*\,\mathrm{d}E=\mathrm{d}S\,.
\end{equation}

In order to study global thermodynamic stability, we now investigate the on-shell Euclidean action of the Euclidean Schwarzschild-AdS solution. This quantity, in turn, is proportional to the Helmoltz free energy,
\begin{equation}
I = \beta_*\,F = \beta_* E-S=\frac{r_+^{d-2} \Omega _{d-2}}{4 G_d \left[(d-3) \ell^2+(d-1) r_+^2\right]}(\ell^2-r_+^2)\, ,
\end{equation}
where we have used the fact that
It follows that when $r_+\geq \ell$ the Euclidean Schwarszschild-AdS solution has lower Euclidean action than thermal AdS and is thus the preferred phase \cite{Hawking:1982dh}. This corresponds to
\begin{equation}
\beta_*\leq \beta^{\mathrm{HP}}\equiv\frac{2\pi}{d-2}\,\ell\,.
\end{equation}
and marks the onset of the so-called Hawking-Page transition.

Note in particular that $\beta^{\star}>\beta^{\mathrm{HP}}$, so that when the Hawking page transition takes place, the black hole is large and has positive specific heat. It is also not a coincidence that $\beta^{star}$ matches the maximum value of $\beta_*$ (at which the large and small Euclidean Schwarzschild-AdS black holes coincide).
In the flat space limit, when $\ell\to+\infty$, the Euclidean Schwarzschild black hole always has negative specific heat and the Hawking-Page transition never occurs.

\subsection{Inside a spherical box}
\label{sec:addcav}

We now analyze the same issues for Schwarzschild-AdS black holes inside a spherical reflecting cavity on which the induced metric is held fixed.  We take the area-radius of the cavity to be $r=r_0$. As in flat space \cite{Hawking:1976de,Gibbons:1976pt,Hawking:1979ig,Page:1981an,Brown:1989fa}, the cavity has an effect on the thermodynamics of the system. In particular, Eq.~(\ref{eq:cav}) expressing the relation between $\beta_*$ and $r_+$ is now affected by the red-shift of the cavity walls
\begin{equation}
\beta_* = 4\pi\frac{\sqrt{f(r_0)}}{\left|f^\prime(r_+)\right|}\,.
\end{equation}

For each value of $\beta_*$, there are again two distinct black solutions, corresponding to small and large black holes. We will refrain from present explicit expressions in terms of $r$, as For numerical purposes it is more useful for introduce a compact coordinate $y$, so that $y=0$ is the Euclidean horizon and $y=1$ is the location of the cavity. In particular, we define \cite{Headrick:2006ti}
\begin{equation}
r=\frac{r_+}{1-\left(1-\frac{r_+}{r_0}\right)y}, \ {\rm so \ that} \  y = \frac{r_0}{r}\frac{r-r_+}{r_0-r_+}\,,
\label{eq:compact}
\end{equation}
with $y\in(0,1)$. It is also useful to introduce dimensionless quantities
\begin{equation}
\label{eq:hlambda}
y_0\equiv \frac{r_0}{r_+}\quad \text{and}\quad y_+\equiv \frac{r_+}{\ell}\,,
\end{equation}
so that we recover the vanishing cosmological constant case for $y_+\to0$.

We can now use these dimensionless quantities to investigate the boundary between large and small black holes. This is also the boundary that we expect to divide regions of moduli space with positive and negative values of $\tilde{\lambda}$. The expression marking this boundary reads
\begin{multline}
y_+^{\star}(y_0)=\Bigg\{\frac{d-3}{2 (d-1)}-\frac{3 d-11+2 y_0^{d-3}}{2 \left(d-3+2 y_0^{d-1}\right)}+
\\
\sqrt{\left[\frac{d-3}{2 (d-1)}-\frac{3 d-11+2 y_0^{d-3}}{2 \left(d-3+2 y_0^{d-1}\right)}\right]^2-\frac{(d-3)}{(d-1) y_0^2} \left[\frac{d-3+(d-1) y_0^2}{d-3+2 y_0^{d-1}}-1\right]}\Bigg\}^{1/2}\,.
\label{eq:crazyl}
\end{multline}
Note that when $y_0\to+\infty$ we recover Eq.~(\ref{eq:trans}).  Furthermore, inside a cavity, even when $\ell\to+\infty$ (and thus $y_+\to0$), there are distinct large and small black hole branches which merge when
\begin{equation}
y_0^{\star}=\left(\frac{d-1}{2}\right)^{\frac{1}{d-3}}\,.
\end{equation}
The above result has been also derived in \cite{Gregory:2001bd}, where a detailed analysis of the negative mode of a Schwarzschild black hole inside a cavity has been performed across a number of spacetime dimensions.

For $d=4$, we recover the well known result $y_0^{\star}=3/2$ \cite{Hawking:1979ig}. We shall see that our numerical results in the canonical ensemble are in perfect agreement with the fact that the solutions are thermodynamically stable iff $y_+ < y_+^\star$. This is certainly not a surprise given the by now very well established relation between the existence of negative modes with grand canonical ensemble boundary conditions and local thermodynamic stability \cite{Reall:2001ag,Dias:2010eu,Hollands:2012sf}.

\bibliographystyle{jhep}
	\cleardoublepage

\renewcommand*{\bibname}{References}

\bibliography{negative}

\providecommand{\href}[2]{#2}\begingroup\raggedright\begin{thebibliography}{10}

\bibitem{Gibbons:1976ue}
G.~W. Gibbons and S.~W. Hawking, \emph{{Action Integrals and Partition
  Functions in Quantum Gravity}},
  \href{http://dx.doi.org/10.1103/PhysRevD.15.2752}{\emph{Phys. Rev. D}
  {\bfseries 15} (1977) 2752--2756}.

\bibitem{Hartle:2020glw}
J.~B. Hartle and K.~Schleich, \emph{{The Conformal Rotation in Linearised
  Gravity}},  in \emph{Quantum Field Theory and Quantum Statistics} (C.~J.~I.
  I.~A.~Batalin and G.~A. Vilkovisky, eds.), pp.~{67--87}, 4, 1987.
\newblock \href{https://arxiv.org/abs/2004.06635}{{\ttfamily 2004.06635}}.

\bibitem{Schleich:1987fm}
K.~Schleich, \emph{{Conformal Rotation in Perturbative Gravity}},
  \href{http://dx.doi.org/10.1103/PhysRevD.36.2342}{\emph{Phys. Rev. D}
  {\bfseries 36} (1987) 2342--2363}.

\bibitem{Mazur:1989by}
P.~O. Mazur and E.~Mottola, \emph{{The Gravitational Measure, Solution of the
  Conformal Factor Problem and Stability of the Ground State of Quantum
  Gravity}}, \href{http://dx.doi.org/10.1016/0550-3213(90)90268-I}{\emph{Nucl.
  Phys. B} {\bfseries 341} (1990) 187--212}.

\bibitem{Marolf:1996gb}
D.~Marolf, \emph{{Path integrals and instantons in quantum gravity:
  Minisuperspace models}},
  \href{http://dx.doi.org/10.1103/PhysRevD.53.6979}{\emph{Phys. Rev. D}
  {\bfseries 53} (1996) 6979--6990},
  [\href{https://arxiv.org/abs/gr-qc/9602019}{{\ttfamily gr-qc/9602019}}].

\bibitem{Dasgupta:2001ue}
A.~Dasgupta and R.~Loll, \emph{{A Proper time cure for the conformal sickness
  in quantum gravity}},
  \href{http://dx.doi.org/10.1016/S0550-3213(01)00227-9}{\emph{Nucl. Phys. B}
  {\bfseries 606} (2001) 357--379},
  [\href{https://arxiv.org/abs/hep-th/0103186}{{\ttfamily hep-th/0103186}}].

\bibitem{Ambjorn:2002gr}
J.~Ambjorn, A.~Dasgupta, J.~Jurkiewicz and R.~Loll, \emph{{A Lorentzian cure
  for Euclidean troubles}},
  \href{http://dx.doi.org/10.1016/S0920-5632(01)01903-X}{\emph{Nucl. Phys. B
  Proc. Suppl.} {\bfseries 106} (2002) 977--979},
  [\href{https://arxiv.org/abs/hep-th/0201104}{{\ttfamily hep-th/0201104}}].

\bibitem{Feldbrugge:2017kzv}
J.~Feldbrugge, J.-L. Lehners and N.~Turok, \emph{{Lorentzian Quantum
  Cosmology}}, \href{http://dx.doi.org/10.1103/PhysRevD.95.103508}{\emph{Phys.
  Rev. D} {\bfseries 95} (2017) 103508},
  [\href{https://arxiv.org/abs/1703.02076}{{\ttfamily 1703.02076}}].

\bibitem{Feldbrugge:2017fcc}
J.~Feldbrugge, J.-L. Lehners and N.~Turok, \emph{{No smooth beginning for
  spacetime}},
  \href{http://dx.doi.org/10.1103/PhysRevLett.119.171301}{\emph{Phys. Rev.
  Lett.} {\bfseries 119} (2017) 171301},
  [\href{https://arxiv.org/abs/1705.00192}{{\ttfamily 1705.00192}}].

\bibitem{Feldbrugge:2017mbc}
J.~Feldbrugge, J.-L. Lehners and N.~Turok, \emph{{No rescue for the no boundary
  proposal: Pointers to the future of quantum cosmology}},
  \href{http://dx.doi.org/10.1103/PhysRevD.97.023509}{\emph{Phys. Rev. D}
  {\bfseries 97} (2018) 023509},
  [\href{https://arxiv.org/abs/1708.05104}{{\ttfamily 1708.05104}}].

\bibitem{Gibbons:1978ac}
G.~W. Gibbons, S.~W. Hawking and M.~J. Perry, \emph{{Path Integrals and the
  Indefiniteness of the Gravitational Action}},
  \href{http://dx.doi.org/10.1016/0550-3213(78)90161-X}{\emph{Nucl. Phys. B}
  {\bfseries 138} (1978) 141--150}.

\bibitem{Prestidge:1999uq}
T.~Prestidge, \emph{{Dynamic and thermodynamic stability and negative modes in
  Schwarzschild-anti-de Sitter}},
  \href{http://dx.doi.org/10.1103/PhysRevD.61.084002}{\emph{Phys. Rev. D}
  {\bfseries 61} (2000) 084002},
  [\href{https://arxiv.org/abs/hep-th/9907163}{{\ttfamily hep-th/9907163}}].

\bibitem{Dias:2010eu}
O.~J.~C. Dias, P.~Figueras, R.~Monteiro, H.~S. Reall and J.~E. Santos,
  \emph{{An instability of higher-dimensional rotating black holes}},
  \href{http://dx.doi.org/10.1007/JHEP05(2010)076}{\emph{JHEP} {\bfseries 05}
  (2010) 076}, [\href{https://arxiv.org/abs/1001.4527}{{\ttfamily 1001.4527}}].

\bibitem{Headrick:2006ti}
M.~Headrick and T.~Wiseman, \emph{{Ricci flow and black holes}},
  \href{http://dx.doi.org/10.1088/0264-9381/23/23/006}{\emph{Class. Quant.
  Grav.} {\bfseries 23} (2006) 6683--6708},
  [\href{https://arxiv.org/abs/hep-th/0606086}{{\ttfamily hep-th/0606086}}].

\bibitem{DeWitt:1967yk}
B.~S. DeWitt, \emph{{Quantum Theory of Gravity. 1. The Canonical Theory}},
  \href{http://dx.doi.org/10.1103/PhysRev.160.1113}{\emph{Phys. Rev.}
  {\bfseries 160} (1967) 1113--1148}.

\bibitem{Kol:2006ga}
B.~Kol, \emph{{The Power of Action: The Derivation of the Black Hole Negative
  Mode}}, \href{http://dx.doi.org/10.1103/PhysRevD.77.044039}{\emph{Phys. Rev.
  D} {\bfseries 77} (2008) 044039},
  [\href{https://arxiv.org/abs/hep-th/0608001}{{\ttfamily hep-th/0608001}}].

\bibitem{Monteiro:2008wr}
R.~Monteiro and J.~E. Santos, \emph{{Negative modes and the thermodynamics of
  Reissner-Nordstrom black holes}},
  \href{http://dx.doi.org/10.1103/PhysRevD.79.064006}{\emph{Phys. Rev. D}
  {\bfseries 79} (2009) 064006},
  [\href{https://arxiv.org/abs/0812.1767}{{\ttfamily 0812.1767}}].

\bibitem{Gratton:1999ya}
S.~Gratton and N.~Turok, \emph{{Cosmological perturbations from the no boundary
  Euclidean path integral}},
  \href{http://dx.doi.org/10.1103/PhysRevD.60.123507}{\emph{Phys. Rev. D}
  {\bfseries 60} (1999) 123507},
  [\href{https://arxiv.org/abs/astro-ph/9902265}{{\ttfamily
  astro-ph/9902265}}].

\bibitem{Gratton:2000fj}
S.~Gratton and N.~Turok, \emph{{Homogeneous modes of cosmological instantons}},
  \href{http://dx.doi.org/10.1103/PhysRevD.63.123514}{\emph{Phys. Rev. D}
  {\bfseries 63} (2001) 123514},
  [\href{https://arxiv.org/abs/hep-th/0008235}{{\ttfamily hep-th/0008235}}].

\bibitem{Gratton:2001gw}
S.~Gratton, A.~Lewis and N.~Turok, \emph{{Closed universes from cosmological
  instantons}}, \href{http://dx.doi.org/10.1103/PhysRevD.65.043513}{\emph{Phys.
  Rev. D} {\bfseries 65} (2002) 043513},
  [\href{https://arxiv.org/abs/astro-ph/0111012}{{\ttfamily
  astro-ph/0111012}}].

\bibitem{Anderson2008}
M.~T. Anderson, \emph{On boundary value problems for einstein metrics},
  \href{http://dx.doi.org/10.2140/gt.2008.12.2009}{\emph{Geometry \& Topology}
  {\bfseries 12} (Jul, 2008) 2009–2045}.

\bibitem{Anderson:2007jpe}
M.~T. Anderson, \emph{{Extension of symmetries on Einstein manifolds with
  boundary}},  \href{https://arxiv.org/abs/0704.3373}{{\ttfamily 0704.3373}}.

\bibitem{Witten:2018lgb}
E.~Witten, \emph{{A Note On Boundary Conditions In Euclidean Gravity}},
  \href{https://arxiv.org/abs/1805.11559}{{\ttfamily 1805.11559}}.

\bibitem{Fournodavlos:2020wde}
G.~Fournodavlos and J.~Smulevici, \emph{{The Initial Boundary Value Problem for
  the Einstein Equations with Totally Geodesic Timelike Boundary}},
  \href{http://dx.doi.org/10.1007/s00220-021-04141-8}{\emph{Commun. Math.
  Phys.} {\bfseries 385} (2021) 1615--1653},
  [\href{https://arxiv.org/abs/2006.01498}{{\ttfamily 2006.01498}}].

\bibitem{Fournodavlos:2021eye}
G.~Fournodavlos and J.~Smulevici, \emph{{The initial boundary value problem in
  General Relativity: the umbilic case}},
  \href{https://arxiv.org/abs/2104.08851}{{\ttfamily 2104.08851}}.

\bibitem{Andrade:2015qea}
T.~Andrade, W.~R. Kelly and D.~Marolf, \emph{{Einstein\textendash{}Maxwell
  Dirichlet walls, negative kinetic energies, and the adiabatic approximation
  for extreme black holes}},
  \href{http://dx.doi.org/10.1088/0264-9381/32/19/195017}{\emph{Class. Quant.
  Grav.} {\bfseries 32} (2015) 195017},
  [\href{https://arxiv.org/abs/1503.03915}{{\ttfamily 1503.03915}}].

\bibitem{Adam:2011dn}
A.~Adam, S.~Kitchen and T.~Wiseman, \emph{{A numerical approach to finding
  general stationary vacuum black holes}},
  \href{http://dx.doi.org/10.1088/0264-9381/29/16/165002}{\emph{Class. Quant.
  Grav.} {\bfseries 29} (2012) 165002},
  [\href{https://arxiv.org/abs/1105.6347}{{\ttfamily 1105.6347}}].

\bibitem{Figueras:2011va}
P.~Figueras, J.~Lucietti and T.~Wiseman, \emph{{Ricci solitons, Ricci flow, and
  strongly coupled CFT in the Schwarzschild Unruh or Boulware vacua}},
  \href{http://dx.doi.org/10.1088/0264-9381/28/21/215018}{\emph{Class. Quant.
  Grav.} {\bfseries 28} (2011) 215018},
  [\href{https://arxiv.org/abs/1104.4489}{{\ttfamily 1104.4489}}].

\bibitem{Gibbons:1978ji}
G.~W. Gibbons and M.~J. Perry, \emph{{Quantizing Gravitational Instantons}},
  \href{http://dx.doi.org/10.1016/0550-3213(78)90434-0}{\emph{Nucl. Phys. B}
  {\bfseries 146} (1978) 90--108}.

\bibitem{Gross:1982cv}
D.~J. Gross, M.~J. Perry and L.~G. Yaffe, \emph{{Instability of Flat Space at
  Finite Temperature}},
  \href{http://dx.doi.org/10.1103/PhysRevD.25.330}{\emph{Phys. Rev. D}
  {\bfseries 25} (1982) 330--355}.

\bibitem{Allen:1984bp}
B.~Allen, \emph{{Euclidean Schwarzschild negative mode}},
  \href{http://dx.doi.org/10.1103/PhysRevD.30.1153}{\emph{Phys. Rev. D}
  {\bfseries 30} (1984) 1153--1157}.

\bibitem{Monteiro:2009ke}
R.~Monteiro, M.~J. Perry and J.~E. Santos, \emph{{Semiclassical instabilities
  of Kerr-AdS black holes}},
  \href{http://dx.doi.org/10.1103/PhysRevD.81.024001}{\emph{Phys. Rev. D}
  {\bfseries 81} (2010) 024001},
  [\href{https://arxiv.org/abs/0905.2334}{{\ttfamily 0905.2334}}].

\bibitem{Monteiro:2009tc}
R.~Monteiro, M.~J. Perry and J.~E. Santos, \emph{{Thermodynamic instability of
  rotating black holes}},
  \href{http://dx.doi.org/10.1103/PhysRevD.80.024041}{\emph{Phys. Rev. D}
  {\bfseries 80} (2009) 024041},
  [\href{https://arxiv.org/abs/0903.3256}{{\ttfamily 0903.3256}}].

\bibitem{Dias:2009iu}
O.~J.~C. Dias, P.~Figueras, R.~Monteiro, J.~E. Santos and R.~Emparan,
  \emph{{Instability and new phases of higher-dimensional rotating black
  holes}}, \href{http://dx.doi.org/10.1103/PhysRevD.80.111701}{\emph{Phys. Rev.
  D} {\bfseries 80} (2009) 111701},
  [\href{https://arxiv.org/abs/0907.2248}{{\ttfamily 0907.2248}}].

\bibitem{MicroPaper}
D.~Marolf and J.~E. Santos, \emph{{Stability of the microcanonical ensemble in
  Euclidean Quantum Gravity - in preparation}},  2022.

\bibitem{Kudoh:2006bp}
H.~Kudoh, \emph{{Origin of black string instability}},
  \href{http://dx.doi.org/10.1103/PhysRevD.73.104034}{\emph{Phys. Rev. D}
  {\bfseries 73} (2006) 104034},
  [\href{https://arxiv.org/abs/hep-th/0602001}{{\ttfamily hep-th/0602001}}].

\bibitem{toappear}
D.~Marolf and J.~E. Santos, \emph{{Stability of thermodynamic ensembles with
  rotation in Euclidean Quantum Gravity - in preparation}},  2022.

\bibitem{Dias:2015nua}
O.~J.~C. Dias, J.~E. Santos and B.~Way, \emph{{Numerical Methods for Finding
  Stationary Gravitational Solutions}},
  \href{http://dx.doi.org/10.1088/0264-9381/33/13/133001}{\emph{Class. Quant.
  Grav.} {\bfseries 33} (2016) 133001},
  [\href{https://arxiv.org/abs/1510.02804}{{\ttfamily 1510.02804}}].

\bibitem{Canuto2010SpectralMF}
C.~Canuto, M.~Y. Hussaini, A.~Quarteroni and T.~A. Zang, \emph{Spectral
  Methods: Fundamentals in Single Domains}.
\newblock Springer-Verlag, 2010.

\bibitem{Hubeny:2002xn}
V.~E. Hubeny and M.~Rangamani, \emph{{Unstable horizons}},
  \href{http://dx.doi.org/10.1088/1126-6708/2002/05/027}{\emph{JHEP} {\bfseries
  05} (2002) 027}, [\href{https://arxiv.org/abs/hep-th/0202189}{{\ttfamily
  hep-th/0202189}}].

\bibitem{Asnin:2007rw}
V.~Asnin, D.~Gorbonos, S.~Hadar, B.~Kol, M.~Levi and U.~Miyamoto, \emph{{High
  and Low Dimensions in The Black Hole Negative Mode}},
  \href{http://dx.doi.org/10.1088/0264-9381/24/22/015}{\emph{Class. Quant.
  Grav.} {\bfseries 24} (2007) 5527--5540},
  [\href{https://arxiv.org/abs/0706.1555}{{\ttfamily 0706.1555}}].

\bibitem{Dias:2015pda}
O.~J.~C. Dias, J.~E. Santos and B.~Way, \emph{{Lumpy AdS$_{5}$\texttimes{}
  S$^{5}$ black holes and black belts}},
  \href{http://dx.doi.org/10.1007/JHEP04(2015)060}{\emph{JHEP} {\bfseries 04}
  (2015) 060}, [\href{https://arxiv.org/abs/1501.06574}{{\ttfamily
  1501.06574}}].

\bibitem{Balasubramanian:1999re}
V.~Balasubramanian and P.~Kraus, \emph{{A Stress tensor for Anti-de Sitter
  gravity}}, \href{http://dx.doi.org/10.1007/s002200050764}{\emph{Commun. Math.
  Phys.} {\bfseries 208} (1999) 413--428},
  [\href{https://arxiv.org/abs/hep-th/9902121}{{\ttfamily hep-th/9902121}}].

\bibitem{deHaro:2000vlm}
S.~de~Haro, S.~N. Solodukhin and K.~Skenderis, \emph{{Holographic
  reconstruction of space-time and renormalization in the AdS / CFT
  correspondence}},
  \href{http://dx.doi.org/10.1007/s002200100381}{\emph{Commun. Math. Phys.}
  {\bfseries 217} (2001) 595--622},
  [\href{https://arxiv.org/abs/hep-th/0002230}{{\ttfamily hep-th/0002230}}].

\bibitem{Gibbons:2004ai}
G.~W. Gibbons, M.~J. Perry and C.~N. Pope, \emph{{The First law of
  thermodynamics for Kerr-anti-de Sitter black holes}},
  \href{http://dx.doi.org/10.1088/0264-9381/22/9/002}{\emph{Class. Quant.
  Grav.} {\bfseries 22} (2005) 1503--1526},
  [\href{https://arxiv.org/abs/hep-th/0408217}{{\ttfamily hep-th/0408217}}].

\bibitem{Hawking:1982dh}
S.~W. Hawking and D.~N. Page, \emph{{Thermodynamics of Black Holes in anti-De
  Sitter Space}}, \href{http://dx.doi.org/10.1007/BF01208266}{\emph{Commun.
  Math. Phys.} {\bfseries 87} (1983) 577}.

\bibitem{Hawking:1976de}
S.~W. Hawking, \emph{{Black Holes and Thermodynamics}},
  \href{http://dx.doi.org/10.1103/PhysRevD.13.191}{\emph{Phys. Rev. D}
  {\bfseries 13} (1976) 191--197}.

\bibitem{Gibbons:1976pt}
G.~W. Gibbons and M.~J. Perry, \emph{{Black Holes and Thermal Green's
  Functions}}, \href{http://dx.doi.org/10.1098/rspa.1978.0022}{\emph{Proc. Roy.
  Soc. Lond. A} {\bfseries 358} (1978) 467--494}.

\bibitem{Hawking:1979ig}
S.~W. Hawking and W.~Israel, \emph{{General Relativity}: {An Einstein Centenary
  Survey}}.
\newblock Univ. Pr., Cambridge, UK, 1979.

\bibitem{Page:1981an}
D.~N. Page, \emph{{Black Hole Formation in a Box}},
  \href{http://dx.doi.org/10.1007/BF00759861}{\emph{Gen. Rel. Grav.} {\bfseries
  13} (1981) 1117--1126}.

\bibitem{Brown:1989fa}
J.~D. Brown, G.~L. Comer, E.~A. Martinez, J.~Melmed, B.~F. Whiting and J.~W.
  York, Jr., \emph{{Thermodynamic Ensembles and Gravitation}},
  \href{http://dx.doi.org/10.1088/0264-9381/7/8/020}{\emph{Class. Quant. Grav.}
  {\bfseries 7} (1990) 1433--1444}.

\bibitem{Gregory:2001bd}
J.~P. Gregory and S.~F. Ross, \emph{{Stability and the negative mode for
  Schwarzschild in a finite cavity}},
  \href{http://dx.doi.org/10.1103/PhysRevD.64.124006}{\emph{Phys. Rev. D}
  {\bfseries 64} (2001) 124006},
  [\href{https://arxiv.org/abs/hep-th/0106220}{{\ttfamily hep-th/0106220}}].

\bibitem{Reall:2001ag}
H.~S. Reall, \emph{{Classical and thermodynamic stability of black branes}},
  \href{http://dx.doi.org/10.1103/PhysRevD.64.044005}{\emph{Phys. Rev. D}
  {\bfseries 64} (2001) 044005},
  [\href{https://arxiv.org/abs/hep-th/0104071}{{\ttfamily hep-th/0104071}}].

\bibitem{Hollands:2012sf}
S.~Hollands and R.~M. Wald, \emph{{Stability of Black Holes and Black Branes}},
  \href{http://dx.doi.org/10.1007/s00220-012-1638-1}{\emph{Commun. Math. Phys.}
  {\bfseries 321} (2013) 629--680},
  [\href{https://arxiv.org/abs/1201.0463}{{\ttfamily 1201.0463}}].

\end{thebibliography}\endgroup

\end{document}